\documentclass[a4paper,fleqn,usenatbib]{mnras}

\usepackage{newtxtext,newtxmath}
\usepackage[T1]{fontenc}
\usepackage{ae,aecompl}
\usepackage{pdflscape}
\usepackage{graphicx}	% Including figure files
\usepackage{amsmath}	% Advanced maths commands
\usepackage{amssymb}	% Extra maths symbols
\usepackage{booktabs}
\usepackage{afterpage}
\usepackage{emptypage}
\usepackage{pdflscape}
\usepackage{cleveref}
\usepackage{multirow}
\usepackage{rotating}
\usepackage{threeparttable}
\usepackage[caption = false]{subfig}
\usepackage[colorinlistoftodos]{todonotes}

\newcommand{\D}{\mathrm{d}}
\newcommand{\pt}{\partial}
\newcommand{\EQ}[1] {Equation~(\ref{#1})}
\newcommand{\FIG}[1] {Figure~\ref{#1}}
\newcommand{\TAB}[1] {Table~\ref{#1}}
\newcommand{\SEC}[1] {Section~\ref{#1}}
\newcommand{\APP}[1] {Appendix~\ref{#1}}
\newcommand{\VEC}[1] {{\boldsymbol{{ #1}}}}
\newcommand{\MX}[1] {{\mathbf{{ #1}}}}

\def\ergs{\mathrm{erg}\, \mathrm{s}^{-1}}

\def\gpcyr{\mathrm{Gpc}^{-3}\, \mathrm{yr}^{-1}}

\def\DM{\mathrm{DM}}
\def\DMm{\DM_{\rm MW}}

\def\DMe{\DM_{\rm E}}
\def\DMi{\DM_{\rm IGM}}
\def\DMh{\DM_{\rm host}}

\def\DMs{\DM_{\rm src}}

\def\DMek{\DM_{\mathrm{E},\,k}}
\def\Sk{S_{k}}
\def\wo{w_\mathrm{o}}
\def\wi{w_\mathrm{i}}
\def\wok{w_{\mathrm{o},\,k}}
\def\nf{N_\mathrm{f}}

\def\fz{f_z}
\def\fd{f_{\cal D}}
\def\fs{f_\mathrm{s}}
\def\fe{f_\mathrm{\epsilon}}
\def\fw{f_\mathrm{w}}

\def\sqdeg{\mathrm{deg}^2}
\def\ergs{\mathrm{erg}\, \mathrm{s}^{-1}}

\def\gpcyr{\mathrm{Gpc}^{-3}\, \mathrm{yr}^{-1}}
\def\cmpc{\mathrm{cm}^{-3}\, \mathrm{pc}}

\title[FRB luminosity function -- II]{On the FRB luminosity function -- II. Event rate density}

\author[R. Luo et al.]{
Rui Luo,$^{1,2,3}$\thanks{E-mail: rui.luo@csiro.au} Yunpeng Men,$^{1,2}$ Kejia Lee,$^{1,4}$\thanks{E-mail: kjlee@pku.edu.cn} Weiyang Wang,$^{5,6}$ D.~R. Lorimer\,$^{7,8}$ \newauthor
\, and Bing Zhang\,$^{9}$ \\
\\
$^{1}$Kavli Institute for Astronomy and Astrophysics, Peking University, 
Beijing 100871, China \\
$^{2}$Department of Astronomy, School of Physics, Peking University, Beijing 
100871, China\\
$^{3}$CSIRO Astronomy and Space Science, Australia Telescope National Facility, Box 76, Epping, NSW 1710, Australia \\
$^{4}$National Astronomical Observatories, Chinese Academy of Sciences, 
Beijing 100012, China\\
$^{5}$Key Laboratory for Computational Astrophysics, National Astronomical Observatories, Chinese Academy of Sciences, 20A Datun Road, Beijing 100012, China\\
$^{6}$School of Astronomy and Space Sciences, University of Chinese Academy of Sciences, Beijing 100049, China\\
$^{7}$Department of Physics and Astronomy, West Virginia University, Morgantown, 
WV 26506, USA \\
$^{8}$Center for Gravitational Waves and Cosmology, West Virginia University, 
Chestnut Ridge Research Building, Morgantown, WV 26505, USA \\
$^{9}$Department of Physics and Astronomy, University of Nevada, Las Vegas, NV 89154, USA
}
\date{Accepted 2020 March 9. Received 2020 March 8 in original form 2019 September 14}

\pubyear{2020}

\begin{document}
\label{firstpage}
\pagerange{\pageref{firstpage}--\pageref{lastpage}}
\maketitle

% Abstract of the paper
\begin{abstract}
The luminosity function of Fast Radio Bursts (FRBs), defined as the event rate 
per unit cosmic co-moving volume per unit luminosity, may help to 
reveal the possible origins of FRBs and design the optimal searching strategy. 
With the Bayesian modelling, we measure the FRB luminosity function using 46 known FRBs. Our Bayesian framework self-consistently models the
selection effects, including the survey sensitivity, the telescope beam response, and the
electron distributions from Milky Way / the host galaxy / local environment of FRBs. Different from the previous companion paper, we pay attention to the FRB event rate density and model the event counts of FRB surveys based on the Poisson statistics. 
Assuming a Schechter luminosity function form, we infer (at the 95\% confidence level)
that the characteristic FRB event rate density at the upper cut-off luminosity $L^*=2.9_{-1.7}^{+11.9}\times10^{44}\,\ergs$ is
$\phi^*=339_{-313}^{+1074}\,\gpcyr$, the power-law index is
$\alpha=-1.79_{-0.35}^{+0.31}$, 
and the lower cut-off luminosity is 
$L_0\le9.1\times10^{41}\,\ergs$. The event rate density of FRBs is found to 
be $3.5_{-2.4}^{+5.7}\times10^4\,\gpcyr$ above $10^{42}\,\ergs$, 
$5.0_{-2.3}^{+3.2}\times10^3\,\gpcyr$ above $10^{43}\,\ergs$ , and 
$3.7_{-2.0}^{+3.5}\times10^2\,\gpcyr$ above $10^{44}\,\ergs$. As a result, we find that, for searches conducted at 1.4~GHz, the optimal diameter of single-dish radio telescopes to detect FRBs is 30--40 m. The possible astrophysical implications of the measured event rate density are also discussed in the current paper. \end{abstract}

% Select between one and six entries from the list of approved keywords.
% Don't make up new ones.
\begin{keywords} stars: luminosity function, mass function -- methods: statistical -- methods: data analysis 
\end{keywords}
%%%%%%%%%%%%%%%%%%%%%%%%%%%%%%%%%%%%%%%%%%%%%%%%%%

%%%%%%%%%%%%%%%%% BODY OF PAPER %%%%%%%%%%%%%%%%%%

\section{Introduction}

The origin of Fast Radio Bursts (FRBs) is still unknown, in spite of the fact that the initial 
discovery was made a decade ago \citep{Lorimer07Sci}. Observationally, FRBs are 
radio flashes that last for a few milliseconds and exhibit prominent dispersion 
with peak flux densities ranging from $\sim$0.05 Jy to $\sim$150 Jy \citep[see FRB catalogue, FRBCAT\footnote{http://frbcat.org/};][]{Petroff16PASA}.
Due to the limited beam size of traditional single-dish radio telescopes, the detection rate
was typically low \citep{Lawrence17AJ} and the total number of FRBs was around 30 by 2018. 
The situation changed dramatically in recent years, thanks to the deployment
of telescope arrays with large fields of view (FoV),
e.g. ASKAP with $\mathrm{FoV}=160\,\sqdeg$ \citep{Bannister17ApJ} 
and CHIME with $\mathrm{FoV}>200\,\sqdeg$ \citep{CHIME18ApJ}. Those new facilities have greatly boosted the annual detection rate of FRBs. 
Up to 2020, the number of verified FRB sources has been over 100.

With the growing number of detected FRBs, one can perform statistical 
studies of the FRB population. The studies carried out so far
cover the investigations for source count---flux 
($\log N(\ge F)-\log F$) relation \citep{Vedantham16ApJ, ME18MN, Golpayegani19MN}, $V/V_{\rm max}$ test 
 \citep{Oppermann16MN, Locatelli19A&A}, FRB event rate 
 distribution \citep{Bera16MN, Lawrence17AJ},
 repetition properties \citep{Connor16MNa, Connor18ApJ, Caleb19MN}, and 
 volumetric event rate \citep{Deng19JHEAp, James19MN, L&P19ApJ}.  All 
the above analyses can be, in principal, derived from the FRB luminosity 
function (LF), i.e.~the event rate per unit co-moving volume per unit luminosity 
scale. The LF characterises the basic properties of FRB population,
i.e.~spatial number density, occurrence frequency, and luminosity distribution.  
If the LF is known, one can directly compute the above mentioned statistics and 
compare them to observations. Furthermore, one can calculate the detection rate of telescopes by integrating the LF above the telescope thresholds. This can be used to design the FRB searching plan in the future. The FRB event rate densities at different luminosity ranges may provide hints in investigating their origins and possible progenitors \citep[e.g.][]{Platts19PhR}.

In our previous work \citep[][; hereafter L18]{Luo18MN}, we measured the normalised FRB 
LF (i.e., probability density of FRBs per unit luminosity interval) using 
a sample of 33 FRBs with Bayesian method. 
L18 investigated the 
relative power output of FRB population in different luminosity bins, but did not address 
how frequently FRB events occur per unit time per unit volume in different luminosity ranges. In the current paper, we aims to measure the \emph{true} LF,
i.e., number of FRBs per co-moving volume per unit time per luminosity scale. 

The major differences between this paper and L18 are:
\begin{itemize}
\item L18 did not take into account the exact survey information, such as survey time, field of view, detection number. Those data are used in this paper.
\item No modelling for event occurrence was carried out in L18, whereas we consider both a random process of event counts and the FRB duration distribution in this paper.
\item The sample used in this paper, among which there are many ASKAP detections, is larger than that of L18. 
\end{itemize}

The structure of this paper is organised as follows. In \SEC{sec:bayes},
we extend our previous model to include the epochs of FRB events using
the Poisson Process. The verification of the Bayesian inference algorithm
using the mock data is also made there. The results of the inference
to the observed FRB sample is shown in \SEC{sec:res}. Relevant discussion and 
conclusions are made in \SEC{sec:dis}.

\section{Bayesian Framework}
\label{sec:bayes}
As described in L18, throughout this paper we make extensive use of
Bayes' theorem, which relates the probability density functions (PDFs) involving
data ($\bf X$) to parameters ($\bf \Theta$) as follows,
\begin{equation} P({\bf
\Theta|X})=\frac{P({\bf \Theta})P({\bf X|\Theta})}{P({\bf X})}\,,
\label{eq:bayest}
\end{equation}
where the symbol $|$ denotes a conditional probability.
The \emph{likelihood} function, $P({\bf X|\Theta})$, is the PDF of the data given the model parameters. $P(\bf \Theta|X)$ is the \emph{posterior} PDF, i.e. the PDF of the parameters given the data set. The Bayesian \emph{evidence} is defined as
\begin{equation}
	P({\bf X})=\int P({\bf \Theta})P({\bf X|\Theta})\, \D\VEC{\Theta}\,.
	\label{eq:evid}
\end{equation}
Finally, the \emph{prior} PDF $P(\bf \Theta)$ describes {\it a priori} information about the model. 

\subsection{Observations}

For the application of Bayesian inference in the current work, 
the data $\VEC{X}$ are the observables from FRB observations, and
the parameters $\VEC{\Theta}$ are the theoretical parameters we
want to measure. The observables used in the current work are, peak flux density 
($S_{\rm peak}$), extra-galactic dispersion measure ($\DMe$), pulse width
($w$), number of FRB detected in a given survey ($N$), time length
of survey ($t_{\rm svy}$), and FoV of survey ($\Omega$). Here $\DMe$ is obtained
by subtracting the Galactic contribution using the YMW16 model
\citep{YMW16}. With the exception of $t_{\rm svy}$, $N$ and $\Omega$, which are new 
additions, these parameters had been discussed in L18. 

The data we use come from a variety of pulsar surveys carried out around 1~GHz:
the Parkes Magellanic Cloud Pulsar Survey (\citep[PKS-MC;][]{Lorimer07Sci, 
Zhang19MN}; the High Time Resolution Universe \citep[HTRU;][]{Thornton13Sci, 
Champion16MN, Ravi15ApJ, Petroff15a, Petroff17MN}; the Survey for Pulsars and 
Extragalactic Radio Bursts \citep[SUPERB;][]{Keane16Nat, Ravi16Sci, Bhandari18MN}, 
Pulsar Arecibo L-band Feed Array \citep[PALFA;][]{Spitler14ApJ}; Green Bank 
Telescope Intensity Mapping \citep[GBTIM;][]{Masui15Nat}; UTMOST Southern Sky 
\citep[UTMOST-SS;][]{Caleb17MN}) and the Commensal Real-time ASKAP Fast Transients 
\citep[CRAFT;][]{Shannon18Nat}. We do not include the CHIME discoveries, since we 
focus on the FRB LF whose radio emission is around 1~GHz. To include 400~MHz to 800~MHz 
band of CHIME, the detailed modelling of spectrum is required, which we defer
to a future paper. The properties of these 7 surveys and  
46 FRBs related to the current work are listed in \TAB{tab:svys} and \TAB{tab:frbs}, 
respectively. 

We include the repeating FRB~121102 in our sample. Considering we are studying on the 
FRB occurrence rate in this paper, there is a subtle difference between using the 
initial discovery and later follow-ups. For the initial discovery, one did not 
know the FRB position, FoV, the telescope gain and the observing duration that determine 
the searching volume. By contrast, for later follow-up observations, one already knew the source position so that the observers can design the telescope gain (by choosing the right telescope) and the observation time to monitor the target.
For the purpose of deriving the \emph{true} FRB LF, we \emph{only} use the information of initial discovery to avoid complicated selection effects introduced by follow-up observations. An underlying assumption is 
that we regard the repeating and non-repeating FRBs as a uniform population.

\begin{table*}
\caption{The known FRB surveys}
\centering
\begin{threeparttable}
\begin{tabular}{cccccccccc}
\hline
\hline
Survey & $N_{\rm FRB}$\tnote{a} & $\Omega$\tnote{b} & $t_{\rm svy}$\tnote{c} & $G$\tnote{d} & $T_{\rm sys}$\tnote{e} & BW\tnote{f} & S/N$_{0}$\tnote{g} & $N_\mathrm{p}$\tnote{h} & Ref.\tnote{i}\\
 & & ($\sqdeg$) & (hr) & (K/Jy) & (K) & (MHz) & & &\\
\hline
PKS-MC & 2 & 0.556 & 490.5 & 0.69 & 28 & 288 & 7 & 2 & [1] \\
HTRU\&SUPERB & 19 & 0.556 & 7357 & 0.69 & 28 & 338 & 10 & 2 & [2] \\
PALFA & 1 & 0.024 & 11503.3 & 0.7\tnote{j} & 30 & 322 & 7 & 2 & [1]\\
GBTIM & 1 & 0.055 & 660 & 2.0 & 25 & 200 & 8 & 2 & [3] \\
UTMOST-SS & 3 & 9 & 4320 & 3.0 & 400 & 16 & 10 & 1 & [4] \\
CRAFT & 20 & 160 & 3187.5 & 0.05 & 100 & 336 & 10 & 2 & [5] \\
%UTMOST-2D & 6 & 9 & 8256 & 1.7 & 330 & 31.25 & 9 & 1 & [6] \\
\hline
\end{tabular}
\begin{tablenotes}
\footnotesize
\item (a) Detection number. (b) Field of view in units of $\sqdeg$. (c) Survey time in units of hour. (d) Telescope gain in units of K/Jy. (e) System temperature in units of K. (f) Bandwidth in units of MHz. (g) Threshold of signal-to-noise ratio. (h) Polarization channel number.
\item (i) The references are: [1] \cite{Lawrence17AJ}; [2] \cite{Bhandari18MN}; [3] \cite{Connor16MNb}; [4] \cite{Caleb17MN}; [5] \cite{Shannon18Nat}; %[6] \cite{Farah19arXiv}.
\item (j) The Arecibo FRB (FRB~121102) was probably detected in the sidelobe of 7-beam receiver, the gain of sidelobe is about 0.7 K/Jy \citep{Spitler14ApJ}. 
\end{tablenotes}
\label{tab:svys}
\end{threeparttable}
\end{table*}

\begin{table*}
\caption{The parameters of FRB sample we use}
\centering
\begin{threeparttable}
\begin{tabular}{ccccccccc}
\hline
\hline
FRB & $S_\mathrm{peak}$\tnote{a} & $w$\tnote{b} & $F$\tnote{c} & DM\tnote{d} & $\DMm$\tnote{e} & $\DMm$\tnote{f} & Survey & Ref.\tnote{g} \\
& (Jy) & (ms) & (Jy\,ms) & ($\cmpc$) & ($\cmpc$) & ($\cmpc$) & & \\
\hline
010312 & $0.25$ & $24.3$ & $6.10$ & $1187.0$ & $51.0$ & $67.0$ & PKS-MC & [1]\\
010724 & $30.0$ & $5.0$ & $150.0$ & $375.0$ & $44.6$ & $94.0$ & PKS-MC & [2] \\
090625 & $1.14$ & $1.92$ & $2.19$ & $899.5$ & $31.7$ & $25.5$ & HTRU & [3] \\
110214 & $27.0$ & $1.9$ & $54.0$ & $168.8$ & $31.1$ & $21.1$ & HTRU & [4] \\
110220 & $1.3$ & $5.6$ & $7.28$ & $944.4$ & $34.8$ & $24.1$ & HTRU & [5] \\
110523 & $0.6$ & $1.73$ & $1.04$ & $623.3$ & $43.5$ & $33.0$ & GBTIM & [6] \\
110626 & $0.4$ & $1.4$ & $0.56$ & $723.0$ & $47.5$ & $33.6$ & HTRU & [5]\\
110703 & $0.5$ & $4.3$ & $2.15$ & $1103.6$ & $32.3$ & $23.1$ & HTRU & [5]\\
120127 & $0.5$ & $1.1$ & $0.55$ & $553.3$ & $31.8$ & $20.6$ & HTRU & [5]\\
121002 & $0.43$ & $5.44$ & $2.34$ & $1629.2$ & $74.3$ & $60.5$ & HTRU & [3]\\
121102 & $0.4$ & $3.0$ & $1.2$ & $557.0$ & $188.0$ & $287.1$ & PALFA & [7] \\
130626 & $0.74$ & $1.98$ & $1.47$ & $952.4$ & $66.9$ & $65.1$ & HTRU & [3] \\
130628 & $1.91$ & $0.64$ & $1.22$ & $469.9$ & $52.6$ & $47.0$ & HTRU & [3] \\
130729 & $0.22$ & $15.61$ & $3.43$ & $861.0$ & $31.0$ & $25.4$ & HTRU & [3] \\
131104 & $1.12$ & $2.08$ & $2.33$ & $779.0$ & $71.1$ & $220.2$ & HTRU & [8]\\
140514 & $0.47$ & $2.8$ & $1.32$ & $562.7$ & $34.9$ & $24.2$ & HTRU & [9] \\
150215 & $0.7$ & $2.8$ & $1.96$ & $1105.6$ & $427.2$ & $296.4$ & HTRU & [10] \\
150418 & $2.2$ & $0.8$ & $1.76$ & $776.2$ & $188.5$ & $325.5$ & SUPERB & [11]\\
150610 & $0.7$ & $2.0$ & $1.3$ & $1593.9$ & $122.0$ & $122.9$ & SUPERB & [12]\\
150807 & $128.0$ & $0.35$ & $44.8$ & $266.5$ & $36.9$ & $25.1$ & SUPERB & [13]\\
151206 & $0.3$ & $3.0$ & $0.9$ & $1909.8$ & $160.0$ & $161.0$ & SUPERB & [12]\\
151230 & $0.42$ & $4.4$ & $1.9$ & $960.4$ & $38.0$ & $37.8$ & SUPERB & [12]\\
160102 & $0.5$ & $3.4$ & $1.8$ & $2596.1$ & $13.0$ & $21.8$ & SUPERB & [12]\\
160317 & $3.0$ & $21.0$ & $63.0$ & $1165.0$ & $319.6$ & $394.6$ & UTMOST-SS & [14]\\
160410 & $7.0$ & $4.0$ & $28.0$ & $278.0$ & $57.7$ & $56.7$ & UTMOST-SS & [14] \\
160608 & $4.3$ & $9.0$ & $38.7$ & $682.0$ & $238.3$ & $310.3$ & UTMOST-SS & [14] \\
170107 & $22.3$ & $2.6$ & $57.98$ & $609.5$ & $35.0$ & $25.2$ & CRAFT & [15]\\
170416 & $19.4$ & $5.0$ & $97.0$ & $523.2$ & $40.0$ & $27.5$ & CRAFT & [15]\\
170428 & $7.7$ & $4.4$ & $34.0$ & $991.7$ & $40.0$ & $27.4$ & CRAFT & [15]\\
170707 & $14.8$ & $3.5$ & $52.0$ & $235.2$ & $36.0$ & $26.9$ & CRAFT & [15]\\
170712 & $37.8$ & $1.4$ & $53.0$ & $312.8$ & $38.0$ & $26.5$ & CRAFT & [15]\\
170906 & $29.6$ & $2.5$ & $74.0$ & $390.3$ & $39.0$ & $26.6$ & CRAFT & [15]\\
171003 & $40.5$ & $2.0$ & $81.0$ & $463.2$ & $40.0$ & $35.4$ & CRAFT & [15]\\
171004 & $22.0$ & $2.0$ & $44.0$ & $304.0$ & $38.0$ & $33.0$ & CRAFT & [15]\\
171019 & $40.5$ & $5.4$ & $219.0$ & $460.8$ & $37.0$ & $26.3$ & CRAFT & [15]\\
171020 & $117.6$ & $1.7$ & $200.0$ & $114.1$ & $38.0$ & $25.8$ & CRAFT & [15]\\
171116 & $19.6$ & $3.2$ & $63.0$ & $618.5$ & $36.0$ & $37.5$ & CRAFT & [15]\\
171213 & $88.6$ & $1.5$ & $133.0$ & $158.6$ & $36.0$ & $33.8$ & CRAFT & [15]\\
171216 & $21.0$ & $1.9$ & $40.0$ & $203.1$ & $37.0$ & $28.7$ & CRAFT & [15]\\
180110 & $128.1$ & $3.2$ & $420.0$ & $715.7$ & $38.0$ & $26.1$ & CRAFT & [15]\\
180119 & $40.7$ & $2.7$ & $110.0$ & $402.7$ & $36.0$ & $37.9$ & CRAFT & [15]\\
180128.0 & $17.5$ & $2.9$ & $51.0$ & $441.4$ & $32.0$ & $26.6$ & CRAFT & [15]\\
180128.2 & $28.7$ & $2.3$ & $66.0$ & $495.9$ & $40.0$ & $28.3$ & CRAFT & [15]\\
180130 & $23.1$ & $4.1$ & $95.0$ & $343.5$ & $39.0$ & $26.1$ & CRAFT & [15]\\
180131 & $22.2$ & $4.5$ & $100.0$ & $657.7$ & $40.0$ & $26.9$ & CRAFT & [15]\\
180212 & $53.0$ & $1.81$ & $96.0$ & $167.5$ & $33.0$ & $27.8$ & CRAFT & [15]\\
\hline
\end{tabular}
\begin{tablenotes}
\footnotesize
\item From left to right, for each FRB, we list (a) peak flux density, $S_{\rm peak}$; (b) pulse width, $w$ (with scattering removed); (c) fluence (pulse engergy), $F$; (d) dispersion measure, DM; (e) Galactic DM calculated by the NE2001 model \citep{CL02}; (f) Galactic DM calculated by the YMW16 model \citep{YMW16}; 
%a code for the pulse survey followed by a reference.
\item (g) The references are: [1] \cite{Zhang19MN}; [2] \cite{Lorimer07Sci}; [3] \cite{Champion16MN}; [4] \cite{Petroff19MN}; [5] \cite{Thornton13Sci}; [6] \cite{Masui15Nat}; [7] \cite{Spitler14ApJ}; [8] \cite{Ravi15ApJ}; [9] \cite{Petroff15a}; [10] \cite{Petroff17MN}; [11] \cite{Keane16Nat}; [12] \cite{Bhandari18MN}; [13] \cite{Ravi16Sci}; [14] \cite{Caleb16MN}; [15] \cite{Shannon18Nat}. 
\end{tablenotes}
\label{tab:frbs}
\end{threeparttable}
\end{table*}

\subsection{Likelihood function}
\label{sec:lik}
The likelihood function used in Bayesian inference is the probability 
distribution function of observables given the model parameters. The likelihood 
we used is an extension of the work in L18, where more details of likelihood 
construction are explained. Here we present a brief introduction and refer the
readers to L18 for further details. Basically, we should make three assumptions 
for the distribution functions of luminosity ($L$), intrinsic duration ($\wi$) 
and random process of FRB events. These are as follows:

{\bf (i)} Following L18, the FRB luminosity is isotropic and its distribution 
follows a Schechter function. Due to the limited FRB number, we neglect the 
cosmic evolution of FRB LF, with the caveat that the star formation rate evolves 
significantly at low redshifts ($z\le1$). There are two reasons why we adopt the Schechter 
function: (1) The Schechter function has a power-law shape and a smooth 
exponential cut-off in the high luminosity end, which makes it easy to compare 
with the other results usually assuming a power-law LF;
(2) The function is widely used for extragalactic objects, such as galaxies, 
quasars, gamma-ray bursts, etc. Similarly, it is rather straightforward to compare the LF of FRBs with the other astronomical objects. 

{\bf (ii)} The FRB intrinsic duration distribution is log-normal in form \citep{Connor19MN} and 
independent of FRB LF. This is a new assumption introduced compared to L18. On one hand, the duration distribution is required, since we now need to compute the detection selection effect of FRBs in a given survey. On the other hand, measuring the duration distribution may be useful for progenitor studies.

{\bf (iii)} The time of arrival for FRBs of a given survey follows the Poisson 
process, a consequence if FRBs are \emph{independent} and \emph{stationary} in a 
unit comoving volume.  

{\bf (iv)} The FRB true position is randomly and uniformly located in the 
telescope beam.

In addition, to derive the joint distribution function,  we make four extra 
assumptions for independence: 

{\bf (v)} The cosmic spatial distribution of FRBs can be treated as homogeneous 
in comoving volume for such a limited sample.

{\bf (vi)} The FRB LF is independent of FRB host galaxies.

{\bf (vii)} The source DM contribution is independent of the host-galaxy DM.

The general Bayesian framework shown in the current paper, however, does not 
depend on the assumption i) to iv), i.e. one can replace the corresponding models 
to perform inference in a similar fashion.

Under the seven assumptions above, we can write the joint distribution function for 
the quantities \{$\log L$, $\log \wi$, $N$, $z$, $\DMh$, $\DMs$, $\log\epsilon$\}, 
which are logarithmic luminosities, logarithmic pulse widths, number of FRBs in a 
given survey, redshifts of FRBs, host galaxg DMs, local DM contributions, and beam 
responses. We refer the readers with interests on this derivation to L18 for the details.
Due to the independence assumed above, the likelihood is the multiplication of 
individual likelihood of each observables, i.e.  
\begin{equation}
\begin{aligned}
	& f(\log L, \log \wi, N, z, \DMh, \DMs, \log\epsilon) \\
	&={}\phi(\log L)\,\fw(\log 
	\wi)\,P(N)\,\fz(z)\,\fd(\DMh|z)\,\fs(\DMs)\,\fe(\log\epsilon)\,.
\end{aligned}
\label{eq:joint}
\end{equation}
Here $\phi(\log L)$ is the LF (see \SEC{sec:lf} for details), 
$\fw(\log \wi)$ is the FRB intrinsic duration distribution (\SEC{sec:fw}), 
$P(N)$ is the Poisson distribution describing the number of events 
(\SEC{sec:poip}), $\fz(z)$ is the FRB cosmic spatial distribution function 
(\SEC{sec:fz}), $\fd(\DMh|z)$ is the host DM distribution function at redshift 
of $z$ (\SEC{sec:fd}), $\fs(\DMs)$ is the source DM distribution function 
(\SEC{sec:fs}) and $\fe(\log\epsilon)$ is the beam response distribution 
function (\SEC{sec:fe}).

We can compute the likelihood function for those observables \{$S_{\rm peak}$, 
$w$, $\DMe$, $N$\} from \EQ{eq:joint}.  The first step is to transform the 
variables using the Jacobian transformation. As shown in \APP{app:margin}, this leads to
\begin{equation}
\begin{aligned}
	& f(\log S, \log \wo, N, \DMe, z, \DMs, \log\epsilon) \\
	& ={} \phi(\log L) \fw(\log \wi) P(N) \fd(\DMh|z) \fz(z) \fs(\DMs) \\
	 & \quad \cdot \fe(\log\epsilon)  (1+z) \,.
\end{aligned}
\end{equation}
Since redshift ($z$), FRB local DM ($\DMs$), and beam response ($\epsilon$) are 
not  direct observables, we marginalise them by integrating 
\EQ{eq:joint}.  The reduced likelihood becomes  
\begin{equation}
\begin{aligned}
\mathcal{L}(\log S, \wo, N, \DMe) &= \prod_{j=1}^M\mathcal{L}_j(N_j)\cdot \prod_{k=1}^{N} f(\log \Sk,\,\log \wok,\,\DMek) \\
&=  \frac{\prod_{j=1}^M (\rho_j\Omega_j t_j)^{N_j}\cdot \exp\left(-\sum_{j=1}^M\rho_j\Omega_j t_j\right)}{\prod_{j=1}^MN_j!} \\
& \quad \cdot\prod_{k=1}^{N} f(\log \Sk,\,\log \wok,\,\DMek)\,,
\end{aligned}
\label{eq:lik}
\end{equation}
where the subscript $j$ refers to the surveys and $k$ refers to the FRBs. $\rho_j$ is the 
event rate of the $j$-th survey, $\Omega_j$ is the FoV of the survey, $t_j$ is its observational 
time, and $N_j$ is the number of FRBs detected in the $j$-th 
survey. The summation of all survey 
detections $N=\sum_{j=1}^{M} N_j$.

The function 
\begin{equation}
\begin{aligned}
	f(\log\Sk,\,\log\wok,\,\DMek)&=\frac{1}{\nf}\int_0^{\infty}
	I(\log L)\, \fw\left(\frac{\wok}{1+z}\right) \fz(z)\, \\ 
	& \quad I(\DMek,\,z)\,(1+z)\,\D z \,,
\end{aligned}
\label{eq:normlik}
\end{equation}
with
\begin{equation}
	I(\log L)=\int \phi(\log L) \fe(\log\epsilon)\, \D\log\epsilon\,.
	\label{eq:alf}
\end{equation}
and
\begin{equation}
I(\DMek,\,z) = \int_0^{\max(\DM_{\rm s})}\fd(\DMh|z) \fs(\DMs)\, \D\DMs \,.
\end{equation}
$\nf$ is the normalisation factor defined as \begin{equation}
\begin{aligned} 
    \nf &= \int_{\log S_{\rm min}(\wo)}^{\infty}\,\D \log S \int \int \int f(\log S, \log \wo, \DMe, z)\, \\
		 & \quad \cdot \D \log \wo \, \D \DMe \, \D z\,,
		 \label{eq:norm}
\end{aligned}
\end{equation}
where the minimum detectable flux density
\begin{equation}
	S_{\mathrm{min}}(\wo)=\frac{\mathrm{S/N}_0\ 
	T_{\mathrm{sys}}}{G\sqrt{N_\mathrm{p}\,\mathrm{BW}\, \wo}}\,.
	\label{eq:smin}
\end{equation}
Here $\mathrm{S/N}_0$ is the signal-to-noise ratio (S/N) threshold for detections in the 
surveys, $T_{\rm sys}$ is the system temperature, $G$ is the telescope gain, 
$N_\mathrm{p}$ is the number of polarization summed, and ${\rm BW}$ is the 
bandwidth.

In the following subsections (from \SEC{sec:lf} to \SEC{sec:poip}), we explain the distribution functions appeard in the above equations individually.

\subsubsection{The form of FRB LF}
\label{sec:lf}
As in L18,
we assume a Schechter LF \citep{Schechter76ApJ} in which the volumetric
event rate density
\begin{equation}
\phi(L)\,\D L=\phi^* 
\left(\frac{L}{L^*}\right)^{\alpha}e^{-\frac{L}{L^*}}\,\D \left(\frac{L}{L^*}\right)\, ,
\label{eq:lf}
\end{equation}
where $\phi^*$ is the characteristic volumetric density of event rate in
units of $\gpcyr$, $\alpha$ is the power-law index, and $L^*$ is the upper 
cut-off luminosity. In logarithmic scale, the form of LF is
\begin{equation}
\phi(\log L)\,\D \log L=\ln10\,\phi^* 
\left(\frac{L}{L^*}\right)^{\alpha+1}e^{-\frac{L}{L^*}}\,\D \log L\,.
\end{equation}

\subsubsection{The intrinsic duration distribution}
\label{sec:fw}
As shown in \EQ{eq:smin}, the selection threshold depends on the FRB duration. We need intrinsic duration distribution to compute the selection biases of FRBs with different observed widths. We assume that the intrinsic FRB width distribution is log-normal in form \citep[see, e.g.,][]{Connor19MN} that
\begin{equation}
	\fw(\log \wi) = \frac{1}{\sqrt{2\pi\sigma_{\rm w}^2}}\exp\left[-\frac{(\log 
	\wi-\mu_{\rm w})^2}{2\sigma_{\rm w}^2}\right],
	\label{eq:wid}
\end{equation}
where $\mu_{\rm w}$ and $\sigma_{\rm w}$ are, respectively, the mean and standard deviation of 
this distribution.

\subsubsection{The cosmic spatial distribution}
\label{sec:fz}
The FRB spatial distribution is assumed to be uniform in co-moving volume, that is 
\begin{equation}
	\fz(z) = \frac{\D V}{\D z}=\frac{cr^2(z)}{H_0E(z)}\,,
	\label{eq:fz}
\end{equation}
where $c$ is the speed of light, $r$ is the comoving distance, $H_0$ is the 
Hubble constant and $E(z)$ is the logarithmic time derivative of the cosmic 
scale factor in a flat $\Lambda$CDM universe, i.e.
\begin{equation}
	E(z)=\sqrt{\Omega_{\rm m}(1+z)^3+\Omega_{\Lambda}} \, .
	\label{eq:funce}
\end{equation}
The cosmology model we use here are from Planck observations, i.e. we take 
$H_{0}=67.8 \, {\rm km\,s^{-1}\,Mpc^{-1}}$, $\Omega_{\rm m}=0.308$ and 
$\Omega_{\Lambda}=0.692$ \citep{Planck16A&A}.

\subsubsection{The DM distribution of host galaxies}
\label{sec:fd}
As shown in L18, the distribution function for $\DMh$ evolves as a function of redshift. L18 showed that
\begin{equation}
	\fd(\DMh|z)= \sqrt{\frac{\mathrm{SFR}(0)}{\mathrm
{SFR}(z)}}\fd\left[\DMh \sqrt{\frac{\mathrm{SFR}(0)}{
	\mathrm{SFR}(z)}}\right]\,,
	\label{eq:dmh}
\end{equation}
where $\mathrm{SFR}(z)$ is the cosmic star formation rate at redshift of $z$.
The function $\fd(\DM_{\rm host,0})$ was computed numerically and approximated 
using multiple Gaussian functions in L18, i.e. we take 
\begin{equation} \fd(\DM_{\rm host,0})\, \D \DM_{\rm host,0}  = \sum_{i=1}^{2} 
	a_{i}
	e^{-\left(\frac{\log \DM_{\rm host,0}-b_i}{c_i}\right)^2}\, \D \DM_{\rm host,0} 
	\,.	\label{eq:dmfit}
\end{equation} As shown in L18, the difference is not substantial when using 
different DM models for host galaxies. To reduce the length of the current paper, 
we merely use the model `ALG(YMW16)" in L18, i.e.  model $\DM_{\rm host,0}$ for all types 
of galaxies with the YMW16 templates. 

\subsubsection{The local DM from FRB source}
\label{sec:fs}
Following L18, due to lack of knowledge on  FRB progenitors, we take the 
least-informative assumption that $\DMs$ follows a uniform distribution in a 
rather wider range from 0 to 50 $\cmpc$, i.e.
\begin{equation}
	\fs(\DMs)=
	\begin{cases}
		1/50 \hspace{3em} 0<\DMs<50\,\cmpc\,, \\
		0 \hspace{10.5em} \textrm{Otherwise.}
	\end{cases}
	\label{eq:dms}
\end{equation}

\subsubsection{The telescope beam response}
\label{sec:fe}
The main beam response distribution function we use here is a Gaussian beam \citep[see, e.g.,][]{BWOptics}. We define the ratio of the observed flux ($S_{\rm obs}$) to the intrinsic flux ($S_{\rm src}$) for an FRB as
\begin{equation}
	\epsilon\equiv\frac{S_{\rm obs}}{S_{\rm src}}=e^{-4 \ln 2 
	\left(\frac{\theta}{\theta_{\rm b}
	}\right)^2}\,,
	\label{eq:gaub}
\end{equation}
where $\theta$ is the angular distance between
the true position of FRB and the beam centre, $\theta_{\rm b}$ is the 
full-width-half-maximum (FWHM) beam size, i.e. $\epsilon=0.5$ for
$\theta=\theta_{\rm b}/2$. 

Adopting a uniform distribution per solid angle for the source position 
inside the FWHM of beam, one will have (L18)
\begin{equation}
	\fe(\log\epsilon)=
	\begin{cases}
		1/\log 2 \hspace{5em} \textrm{Inside}\,, \\
		0 \hspace{6.8em} \textrm{Outside}\,. 
	\end{cases}
	\label{eq:fe}
\end{equation}

\subsubsection{Event counts likelihood}
\label{sec:poip}
In L18, the parameter that contains the units of event rate ($\phi^*$) is canceled, when the marginalisation and normalisation were done (also see \EQ{eq:normlik}). In this work, as we model the random process of event occurrence, $\phi^*$ is kept in the event counts likelihood.

We assume the number of \emph{independent} FRBs events\footnote{Here, 
\emph{independent events} are the events of different FRBs and not including the 
repeating events.} that occur in a certain time interval for a given survey follows the 
Poisson distribution. This requires that the FRB event rate is stationary, and 
the bursts are independent (in a probabilistic sense) with each other.  Assuming 
Poisson statistics, the probability of detecting $N$ FRB events in a 
particular survey,
\begin{equation}
P(N)=\frac{(\rho \Omega t)^N e^{-\rho \Omega t}}{N!},
\label{eq:poi}
\end{equation}
where $\rho$ is the event rate per solid angle, namely surface event rate. $\Omega$ is the field of view and $t$ is the survey time. As shown in \APP{app:rho}, integrating the LF, we find that
\begin{equation}
	\rho = \int_0^{\infty} \frac{1}{1+z}\frac{r(z)^2}{H(z)}\D z  \int_{\log L_{\mathrm{min}}(\wo)}^{\infty} I(\log L)\,\D\log L 
\end{equation}
where $H(z)$ is the Hubble parameter, $r$ is the comoving distance. The low 
limit of integrating the LF, i.e. the minimum observed 
luminosity, is $L_\mathrm{min}(\wo)=\max\left[L_0, L_\mathrm{thre}(\wo)\right]$.
Here, $L_0$ is the intrinsic lower cut-off of LF and 
the survey detection threshold
\begin{equation}
	L_\mathrm{thre}(\wo)\equiv4\pi d_{\rm L}^2 \Delta\nu_0 S_\mathrm{min}\,,
\end{equation}
where $d_{\rm L}$ is the luminosity distance, $\Delta\nu_0$ is the intrinsic spectrum width, and $S_{\rm min}$ is the flux detection threshold as defined in 
\EQ{eq:smin}.

According to \EQ{eq:smin}, $S_{\rm min}$ depends on the observed pulse width $\wo$. 
Hence, the event rate seen by observer (i.e. event rate per solid angle $\rho$)  
depends on $\wo$ as well. As proven in \APP{app:mpoi}, the event rate for all FRBs is 
the summation of event rates of individual ones. As the pulse width is a 
continuous random variable, the summation becomes an integration weighted by the 
width distribution function, i.e.
\begin{equation}
\begin{aligned}
	\rho &= \int_0^{\infty} \frac{1}{1+z}\frac{r(z)^2}{H(z)}\D z \int \fw(\log \wi)\,\D\log \wi \\
	&\quad \cdot \int_{\log L_{\mathrm{min}}(\wo)}^{\infty} I(\log L)\,\D\log L\,.
\end{aligned}
\label{eq:rhoj}
\end{equation}
For the $j$-th survey, the likelihood of detecting $N_j$ FRBs is
\begin{equation}
	\mathcal{L}_{j}(N_j)=\frac{(\rho_j\Omega_jt_j)^{N_j} 
	e^{-\rho_j\Omega_jt_j}}{N_j!}\,,
\end{equation}
where $\rho_j$ is the apparent event rate per solid angle for this survey, which can be calculated using \EQ{eq:rhoj}. 

\subsection{Inference algorithm and its verification}
With the likelihood function \EQ{eq:lik}, we can perform parameter inference 
using standard Bayesian sampling algorithm. Similar to L18, we use the 
\textsc{multinest} algorithm developed by \citet{Feroz09MN} to perform the 
parameter sampling and inference.

We use the simulated data to verify our method. The simulated data is generated 
using Monte-Carlo simulation, where the recipe is summarised as following

1) Generate the FRB luminosities $L$ according to the Schechter function 
(\EQ{eq:lf});

2) Generate the intrinsic FRB pulse widths $\wi$ according to \EQ{eq:wid};

3) Generate FRB redshifts $z$ according to the FRB cosmic spatial distribution 
(\EQ{eq:fz}) and calculate the observed pulse width using $\wo=\wi (1+z)$; 

4) Generate the host galaxy DMs ($\DMh$) at the redshift $z$ using the DM 
distribution function mentioned in \SEC{sec:fd}; 

5) Generate the FRB local DMs according to \EQ{eq:dms};

6) Draw a sample of FRB positions in the beam according to \EQ{eq:gaub}; 

7) Based on the above parameters, calculate the observed FRB flux densities and 
extragalactic DMs; 

8) Select the FRBs above the flux threshold $S_{\rm min}$ for detections. 

9) Simulate the FRB arrival times according to the Poisson distribution.

We perform two simulations with two sets of input parameters. Each simulation 
produces 100 fake FRBs.
The survey information is listed in \TAB{tab:mocsvy}. We then apply our 
parameter inference to the simulated data sets to see if the input parameters 
are recovered. The comparison between the input parameters and inferred
parameters is shown in \APP{app:post}, where the posterior distributions of the 
mock data are in \FIG{fig:post_mock}, the input and inferred parameters are 
listed in \TAB{tab:mocres}. The results show that our method does indeed correctly 
recover the parameters used to simulate the mock data sets, giving us confidence for the applications in the real data set, which is presented in the next section.

\section{Results}
\label{sec:res}
\subsection{Measurements of FRB LF}
\begin{figure*}
	\centering
	\includegraphics[width=\textwidth]{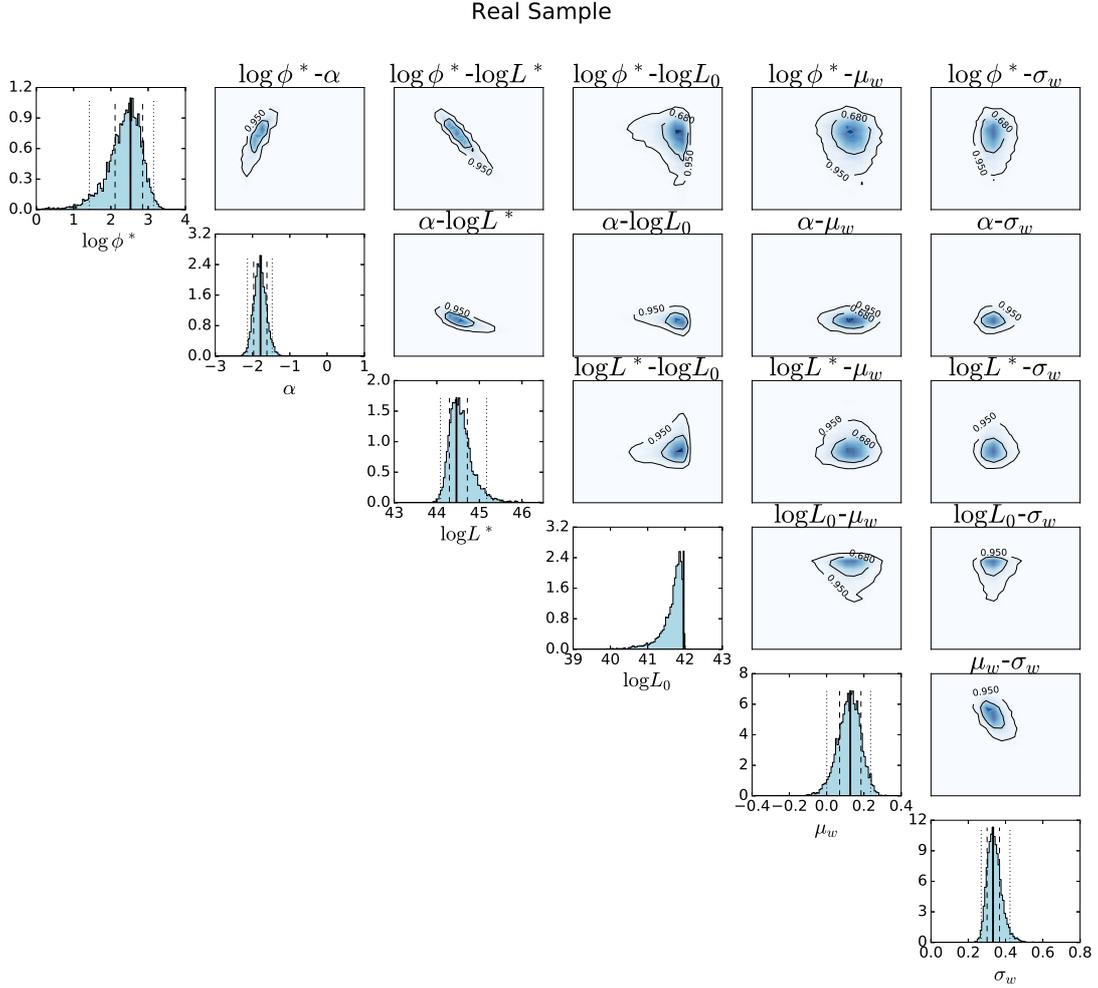}
	\caption{Posterior distributions showing the inferred parameters of the FRB LF. In this ``corner plot'', the 
	diagonal histograms are the marginalised one-dimensional posterior 
	distributions for all the parameters. For the parameters except $\log L_0$, 
	the solid lines denote the most probable measured values, the dashed lines and 
	dotted lines denote 68\% and 95\% confidence interval, respectively. For $\log L_0$, due to 
	the limited sample size, we do not measure the value and the solid line 
	represents the upper limit value with 95\% confidence level (C.L.). The off-diagonal contour 
	plots are for the marginalised two-dimensional posteriors, where the parameter pairs are
	indicated in the plot titles. The inner and outer black contours are for 68\% and 
	95\% confidence intervals, respectively.}
	\label{fig:post}
\end{figure*}

We apply our Bayesian methodology to the 46 FRBs from 7 surveys (\TAB{tab:svys} and \TAB{tab:frbs}). The posterior distributions for the model parameters are shown in \FIG{fig:post}.

\begin{figure}
	\centering
	\subfloat[The FRB LF]
	{\label{fig:lf}\includegraphics[width=3.5in]{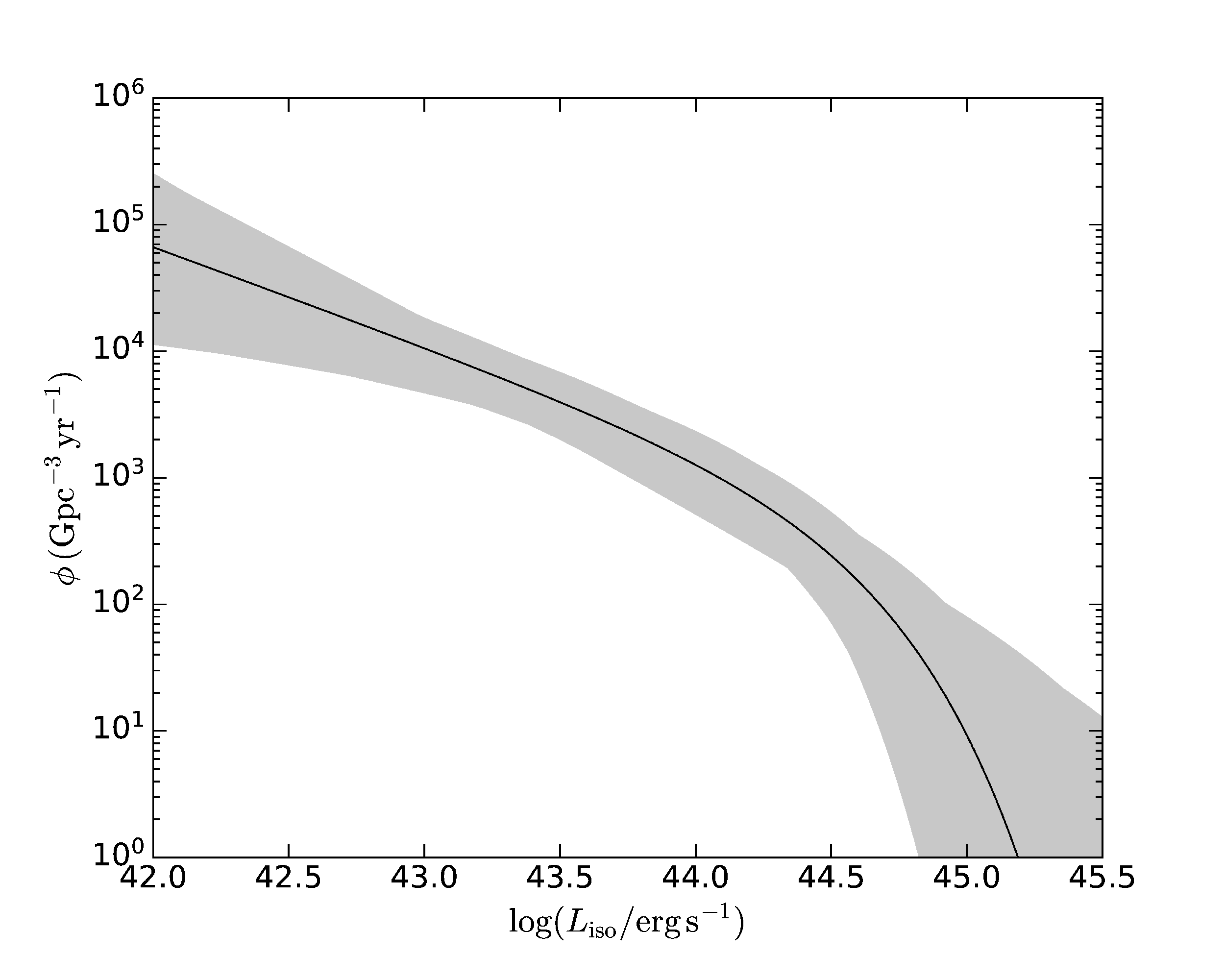}} \\
	\subfloat[The cumulative event rate density of FRBs]
	{\label{fig:ed}\includegraphics[width=3.5in]{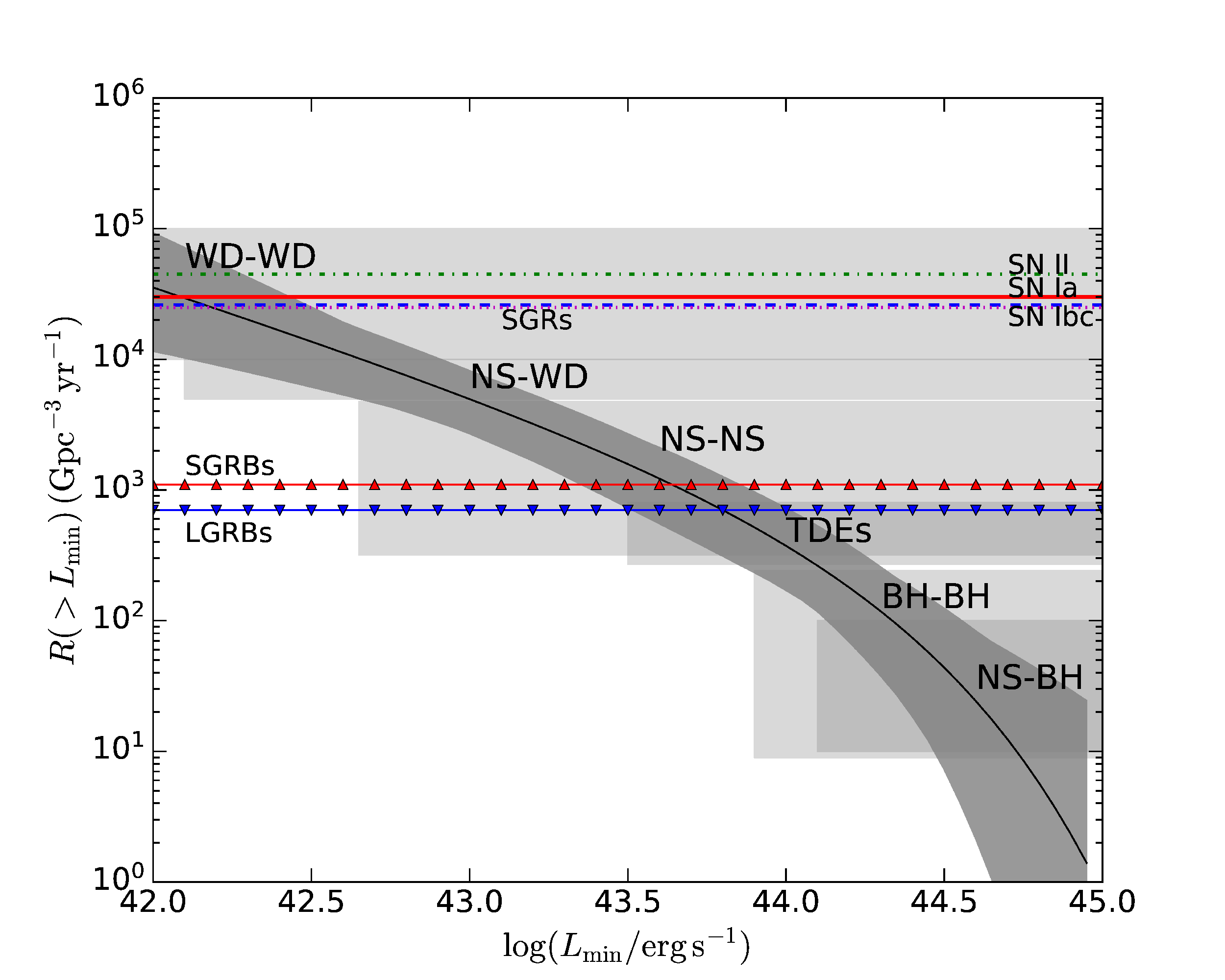}}
	\caption{(a) The x-axis is logarithmic luminosity, the y-axis is the luminosity 
	function defined as event rate per unit co-moving volume per logarithmic luminosity. 
	The black solid curve represents the most probable LF 
	shape with the gray shade indicating 2-$\sigma$ confidence region. 
	(b) The x-axis is the threshold luminosity in logarithmic form, the y-axis is cumulative event rate density above the threshold luminosity. 
	The black solid curve with grey shaded region represents the volumetric event rate of FRB population in 2-$\sigma$ C.L. The merger rate of WD-WD, NS-WD, NS-NS, BH-BH, NS-BH and the TDE rate are marked with six shaded regions. Meanwhile, we also plot the volumetric rate of such transients with six lines, i.e., SN Ia (red solid line), Ibc (blue dashed line), II (green dash-dotted line), SGRs (magenta dotted line), SGRBs (red solid line with up-pointing triangles) and LGRBs (blue solid line with down-pointing triangles).
	}
\end{figure}

The measured LF parameters within 95\% confidence interval are 
$\phi^*=339_{-313}^{+1074}\,\gpcyr$, $\alpha=-1.79_{-0.35}^{+0.31}$ and $\log 
L^*=44.46_{-0.38}^{+0.71}$. We can not measure the value of the lower cut-off 
luminosity, due to the limited size of FRB sample. Its upper limit is $\log 
L_0\le41.96$. The mean value of pulse width distribution is $\mu_w=0.13_{-0.13}^{+0.11}$ and the standard deviation of it is $\sigma_w=0.33_{-0.06}^{+0.09}$. We 
also use the posterior to compute the confidence region of FRB LF, 
where the 2-$\sigma$ region is shown in \FIG{fig:lf}.

To characterize the event rate distribution of the whole FRB population,
we integrate the LF (\EQ{eq:lf}) from a minimum luminosity $L_{\rm min}$ to infinity as follows:
\begin{equation}
	R_{\rm FRB}(>L_{\rm min})=\int_{L_{\rm min}/L^*}^\infty \phi^* \left(\frac{L}{L^*}\right)^\alpha \,\D \left(\frac{L}{L^*}\right) = \phi^* \Gamma\left(\alpha+1, \frac{L_{\rm min}}{L^*}\right)\,,
\end{equation}
where $\Gamma$ is the incomplete \textsc{Gamma} function. The cumulative event rate distribution as a function of $L_{\rm min}$ is shown in \FIG{fig:ed}.

\begin{figure}
\centering
\includegraphics[width=3.5in]{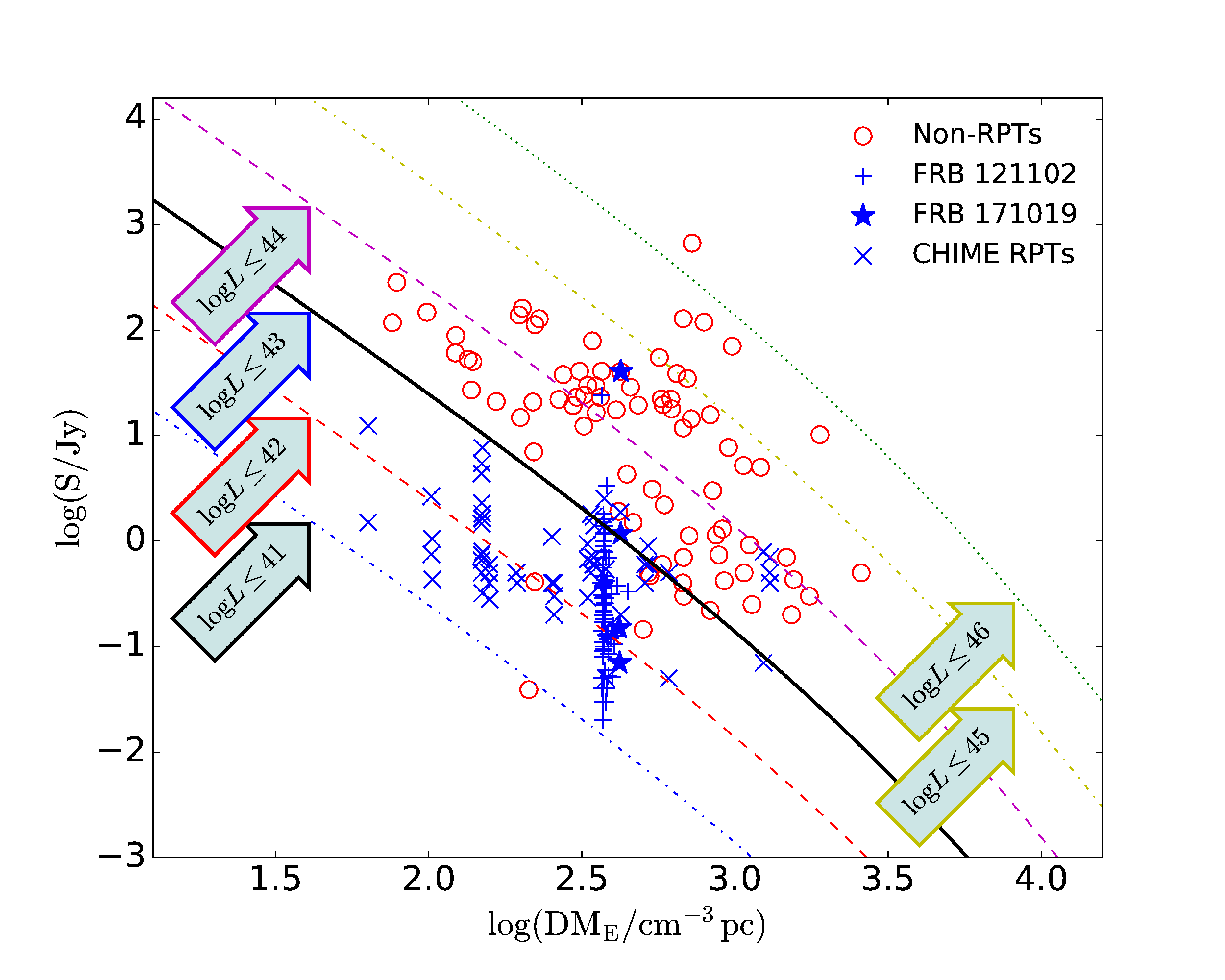}
\caption{Flux-$\DMe$ diagram. The flux density of each FRB is adopted from FRBCAT or corresponding literature, 
$\DMe$ is obtained after removing the Galactic DM using the NE2001 electron model \citep{CL02}
We plot all the FRBs published by January 2020, including repeaters (RPTs) and non-repeaters (Non-RPTs). The red hollow dots denote 
non-repeating FRBs, while repeating FRBs are labelled with blue markers, i.e. FRB 121102 (plus), FRB 171019 (star) and CHIME repeaters (cross).
The six curves that run diagonally are the maximum luminosities assuming that all $\DMe$ comes from 
cosmological contribution without host galaxy or local contribution. For each 
curve, the corresponding arrow indicates the luminosity values. }
\label{fig:fdm}
\end{figure}

Similar to L18, we can also estimate the luminosity of individual FRB without 
marginalisation. The broad luminosity range of all the published FRB sample can 
be visualised clearly on the Flux-$\DMe$ diagram (\FIG{fig:fdm}), which is consistent with results of \cite{Shannon18Nat}. 
In particular, there appears to be a boundary between repeating 
(e.g. FRB~121102 \citep{Spitler16Nat}, FRB~171019 \citep{Kumar19ApJL} and CHIME repeaters \citep{CHIME19Nat02, CHIME19ApJL, CHIME20ApJ}) and non-repeating FRBs around luminosity
$10^{42}-10^{43}\, \ergs$.

\subsection{Detection rate}
\begin{figure}
	\centering
	\includegraphics[width=3.5in]{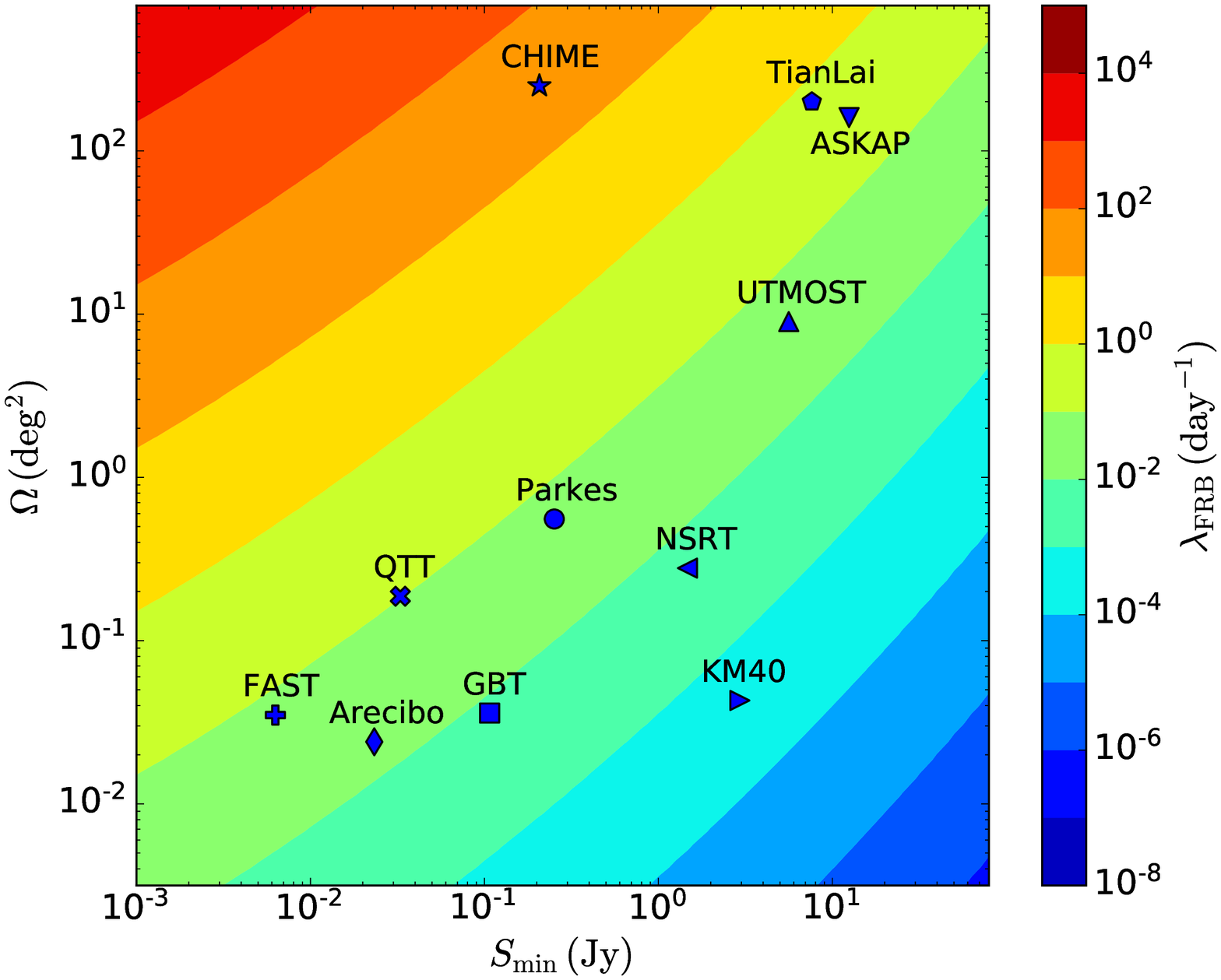}
	\caption{The FRB detection rate map. The x-axis is the detection
	threshold in flux, the y-axis is the field of view, the color bar on the right indicates the 
	apparent detection rate, i.e., expected detection number of FRBs per day. Eleven 
	telescopes are marked in the map. The fields of view and sensitivities come from 
	corresponding references: Parkes \citep{Staveley-Smith96PASA}, Arecibo 
	\citep{Spitler14ApJ}, GBT \citep{Masui15Nat}, UTMOST \citep{Caleb17MN}, ASKAP 
	\citep{Shannon18Nat}, CHIME \citep{CHIME18ApJ}, TianLai \citep{Chen12IJMPS}, 
	FAST\citep{Nan11}, QTT\citep{Wang17}, NSRT and KM40 \citep{Men19MN}. The extension of FoV using 
	multibeam or phase array are considered if available at the site. The 
	signal to noise threshold of valid FRB detections is set as 10 for all of telescopes plotted here.}
	\label{fig:map}
\end{figure}

Using the LF, we can study the apparent detection rate (i.e.,  
the number of detections expected per day) for a given telescope.  The 
detection rate is computed from
\begin{equation}
\begin{aligned}
\lambda(\Omega, S_{\rm min}) &= \frac{\D N_{\mathrm{obs}}(>S_{\rm min})}{\D t_{\mathrm{obs}}}  \\
& = \int_0^{\Omega}\D \Omega \int_0^{\infty} \frac{1}{1+z}\frac{r(z)^2}{H(z)}\,\D z \int \fw(\log \wi)\,\D\log \wi \\
& \quad \cdot \int_{\log L_{\mathrm{min}}(\wo)}^{\infty} I(\log L)\, \D \log L\,
.\\
\end{aligned}
\label{eq:detrate}
\end{equation}
The results are shown in \FIG{fig:map}, where eleven telescopes performing FRB 
searches are also indicated.  

\begin{figure}
	\centering
	\includegraphics[width=3.5in]{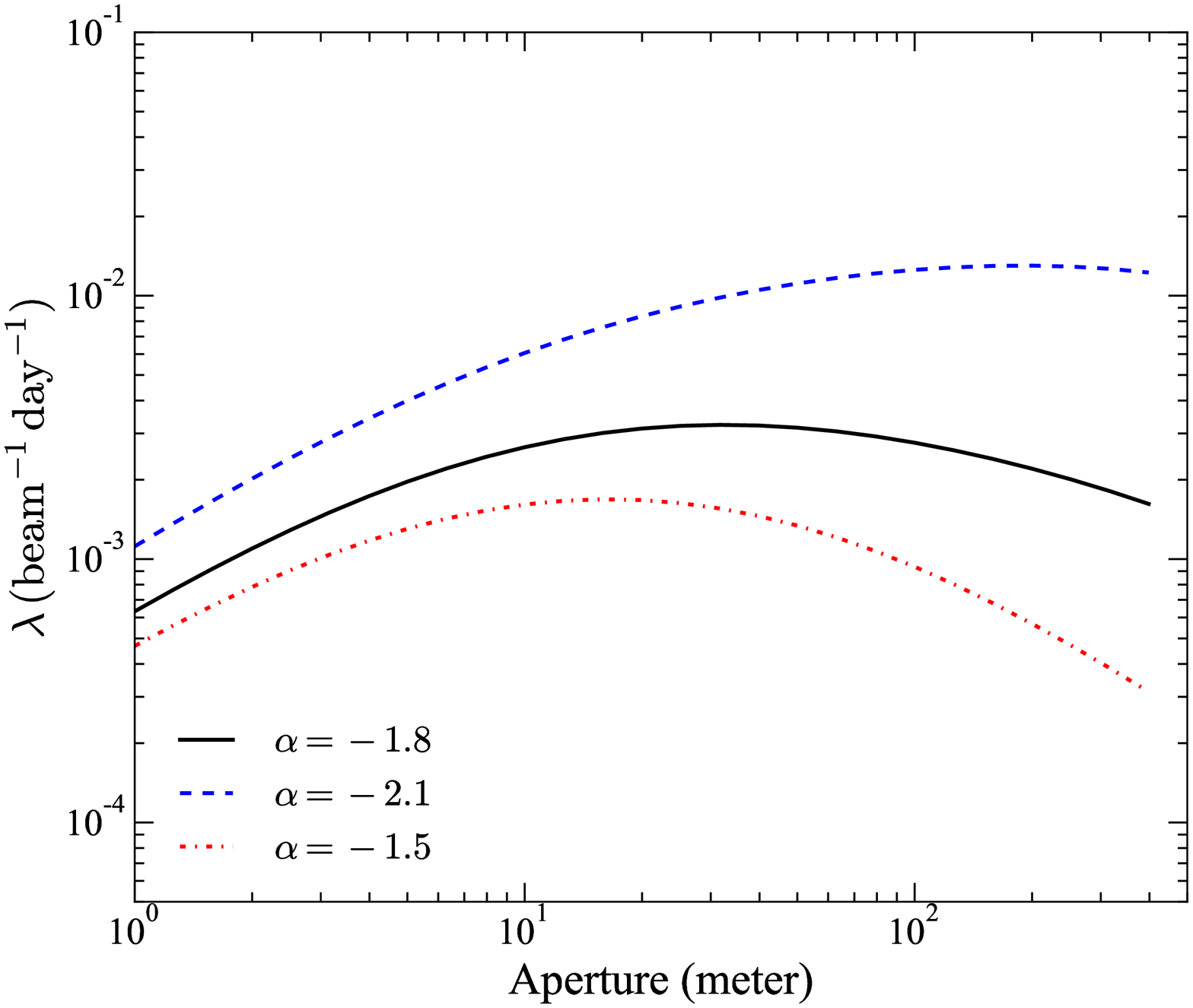}
	\caption{The relation between FRB detection rate and telescope diameter.  In this analysis we have 
	assumed a telescope central frequency of 1.4~GHz, bandwidth 300 MHz, $T_{\rm 
	sys}=25\,{\rm K}$. The detection rate depends on the shape of luminosity 
	function, where three curves with power-law indices of $-1.5$, $-1.8$ and $-2.1$ are 
	plotted, which correspond to our 2-$\sigma$ confidence region of the luminosity 
	power-law index. For the most-likely $\alpha$ value, the optimal telescope diameter is about 30~m.}
	\label{fig:apert}
\end{figure}

For single-dish telescopes without multibeam or phased array feeds, the 
telescope diameter determines both field of view and sensitivity. Thus, for a given 
bandwidth, system temperature, and detection threshold, the detection rate is 
determined only by the diameter of telescope and LF.  As shown in \FIG{fig:apert}, there is an optimal
diameter for the telescope to maximize the detection rate.

Using the detection rate in \EQ{eq:detrate}, we compare the expected detection number with the real detection number of these surveys we used (see \TAB{tab:drate}). These two sets of numbers are consistent within 2-$\sigma$ confidence. 

\begin{table*}
    \caption{Detection numbers check for the FRB surveys used in this work.}
    \centering
    \begin{threeparttable}
    	\begin{tabular}{cccccccc}
    	\hline
    	\hline
    	Survey & PKS-MC & HTRU\&SUPERB & PALFA & GBTIM & UTMOST-SS & CRAFT \\
    	\hline
    	$N_{\rm theo}$\tnote{a} & $1.2_{-1.1}^{+3.9}$ & $14.1_{-13.0}^{+45.2}$ & $1.2_{-1.1}^{+3.9}$ & $0.33_{-0.30}^{+1.05}$ & $5.8_{-5.4}^{+18.7}$ & $28.7_{-26.4}^{+92.0}$ \vspace{0.5em} \\  
    	$N_{\rm real}$\tnote{b} & 2 & 19 & 1 & 1 & 3 & 20 \\
    	\hline
    	\end{tabular}
    	\begin{tablenotes}
    	\footnotesize
    	\item (a) Theoretical detection number with 2-$\sigma$ errors
    	\item (b) The actual  detection number as observed in FRB surveys.
    	\end{tablenotes}
    \end{threeparttable}
    \label{tab:drate}
\end{table*}

\section{Discussion and conclusions}
\label{sec:dis}
\subsection{The measurements}
Our measurements of the power-law index and the upper cut-off luminosity are 
$\alpha=-1.79_{-0.35}^{+0.31}$ and $\log L^*=44.46_{-0.38}^{+0.71}$, 
respectively. For both of the parameters, the values match the previous results of 
L18, who used less number of FRB and obtained $\alpha=-1.57_{-0.37}^{+0.40}$ and 
$\log L^{*}=44.31_{-0.27}^{+0.22}$. The new FRBs from ASKAP sample are brighter, 
and slightly increases the cut-off luminosity. We still have not yet measured 
the intrinsic lower limit of luminosity ($L_0$), due to the missing of sufficient low-luminosity FRBs in the current sample. In the future, such FRBs found by larger 
telescopes should help us to understand and constrain the LF shape and extent at 
the low-luminosity end.

Our inferred event rate density (also called volumetric event rate) agrees with the previous estimations. 
\cite{Cao18ApJ} and \cite{Deng19JHEAp} derived the FRB volumetric event rate as 
$R_{\rm FRB}=(3-6)\times10^4\,\gpcyr$ and $R_{\rm 
FRB}=1.8\pm0.3\times10^4\,\gpcyr$, respectively, using the FRB samples containing 22 and 28 Parkes 
FRBs. Although we have not obtained the intrinsic lower limit of the LF, 
once we integrate the LF above luminosity of $10^{42}\,\ergs$, we can calculate that
$R_{\rm FRB}=3.5_{-2.4}^{+5.7}\times 10^{4}\, \gpcyr$. This is consistent with 
the estimations made by the two previous works. For the other threshold luminosities, 
such as $10^{43}$ and $10^{44}\,\ergs$, our measurements give the event rate densities as $5.0_{-2.3}^{+3.2}\times10^3$ and 
$3.7_{-2.0}^{+3.5}\times10^2\,\gpcyr$, respectively. In particular, our measured volumetric rate distribution is identical with the energy function derived by \cite{L&P19ApJ} from $10^{31}$ to $10^{32}\,\rm erg\,Hz^{-1}$ based on 20 ASKAP FRBs.

Our work focuses on the LF of the sample that mainly consists of non-repeating FRBs. This is 
very different from work by \cite{James19MN}, who derived an upper-limit on 
the repeating FRB density of $\le 27\, {\rm Gpc}^{-3}$ assuming certain 
repeating rate and pulse energy distribution. Although it is not legitimate to 
compare the densities of repeating FRBs with the LF in the current 
paper, the much higher FRB source density in our paper also indicates that the FRB 
121102 is not a typical FRB source as already discussed by 
\cite{James19MN}, see also \cite{Palaniswamy18ApJ, Caleb19MN, Li19ApJ}. Either its burst rate is higher than the majority of FRBs or its burst events are highly correlated in time.

We include no spectral modelling in the current paper and have assumed that the 
FRB spectrum is flat around 1.4~GHz. This is a limitation inherited from the 
current FRB sample, where most of FRBs used in our computation were detected at 
$\sim1.4\,\rm GHz$. The limited frequency coverage also prevents us from  detailed 
modelling of the spectrum. We postpone the spectrum modelling to future work, when a 
wider frequency coverage is available. The flat-spectrum assumption is driven by 
the repeater FRB~121102, for which an apparently flat spectrum with $\sim1\,\rm GHz$ 
bandwidth was observed \citep{Gajjar18ApJ}. 

We have neglected the scattering effects. The reasons for this are twofold: 1) The FRBs used in the 
current paper is not scattering limited, i.e., there is no indication that 
scattering downgrade the S/N by order of magnitude; 2) The uncertainty of our 
LF is still rather large (more than a factor of 10).  Our 
error mainly comes from the limited FRB sample size and the uncertainty of FRB 
distance. The caveat is that the event rate of distant FRBs seen by large 
telescopes (e.g.,  FAST) may be overestimated, since these FRBs might be 
scattering limited.

\subsection{Constraints on the possible origins}

The origins of FRBs are still highly debating and many models have been proposed to explain the phenomena \citep[e.g.][]{Platts19PhR}. The models invoke young pulsars \citep{Cordes16MN,Connor16MN}, relatively old magnetars as observed \citep{PP10,Katz16ApJ}, putative young magnetars born from gamma-ray bursts and superluminous supernovae (SNe) \citep{Murase16,Metzger17ApJ,Wang18},  normal pulsars with external interactions \citep{Zhang17ApJ}, among others. The catastrophic (non-repeating) models include various types of compact star mergers, such as white dwarf (WD) - WD mergers \citep{Kashiyama13ApJ}, neutron star (NS) - NS mergers \citep{Totani2013PASJ,Wang16ApJ}, NS - black hole (BH) mergers \citep{Zhang19ApJL,Dai19ApJL}, and even BH - BH mergers \citep{Zhang16ApJL,Liu16ApJ}, or collapses of supramassive neutron stars into black holes \citep{FR14A&A, Zhang14ApJ}. The expected event rates of these scenarios differ, and by comparing with our results, one can constrain the models.

In \FIG{fig:ed}, we list a few types of burst-like events that are possibly associated with FRBs, including:
1) stellar BH-BH mergers, $R_{\rm BH-BH}=(9-240)\,\gpcyr$ \citep{Abb16b}; 
2) NS-NS mergers, $R_{\rm NS-NS}=1.5_{-1.2}^{+3.2}\times10^3\,\gpcyr$ \citep{Abbott17PRL}; 
3) NS-BH mergers, $R_{\rm NS-BH}=(10-100)\,\gpcyr$ \citep{Map18}; 
4) NS-WD mergers, $R_{\rm NS-WD}=(0.5-1)\times10^4\,\gpcyr$ \citep{Thompson09arXiv}; 
5) WD-WD mergers, $R_{\rm WD-WD}=(10^4-10^5)\,\gpcyr$ \citep{BM12ApJ}; 
6) tidal disruption events (TDEs), $R_{\rm TDEs}=4.8_{-2.1}^{+3.2}\times10^2\,\gpcyr$ \citep{SZL15}; 
7) three types of supernovae (SNe), i.e., SN Ia ($R_{\rm SN Ia}=3\times10^4\,\gpcyr$), Ibc ($R_{\rm SN Ibc}=2.5\times10^4\,\gpcyr$) and II ($R_{\rm SN II}=4.5\times10^4\,\gpcyr$) \citep{Li11MN}; 
8) soft gamma-ray repeaters (SGRs), $R_{\rm SGRs}<2.5\times10^4\,\gpcyr$ \citep{Ofek07ApJ, Kulkarni14ApJ}; 
9) short gamma-ray bursts (SGRBs, beaming-corrected rate), $R_{\rm SGRBs}=1.1\times10^3\,\gpcyr$ \citep{CHP12, SZL15}; 
10) long gamma-ray bursts (LGRBs), $R_{\rm LGRBs}=7\times10^2\,\gpcyr$ \citep{CTP07, SZL15}.
The event rate densities of the transients above and the associated FRB threshold luminosities are summarized in \TAB{tab:asso}.

\begin{table}
    \caption{The possibly associated transients}
	\centering
	\begin{threeparttable}
		\begin{tabular}{cccc}
			\hline
			\hline
			Transient & $R\,(\gpcyr)$\tnote{a} & $\log (L_\mathrm{min}/\ergs)$\tnote{b} & Ref.\tnote{c} \\
			\hline
			BH-BH & $(9-240)$ & $(44.1-44.7)$ & [1] \\
			NS-NS & $1.5_{-1.2}^{+3.2}\times10^3$ & $43.5\pm0.5$ & [2] \\
			NS-BH & $(10-100)$ & $(44.3-44.7)$ & [3] \\
			NS-WD & $(0.5-1)\times10^4$ & $(42.7-43.0)$ & [4] \\
			WD-WD & $(10^4-10^5)$ & $(41.4-42.7)$ & [5] \\
			TDEs & $4.8_{-2.1}^{+3.2}\times10^2$ & $43.9\pm0.2$ & [6] \\
			SN Ia & $3\times10^4$ & $42.1$ & [7] \\
			SN Ibc & $2.5\times10^4$ & $42.2$ & [7] \\
			SN II & $4.5 \times10^4$ & $41.9$ & [7] \\
			SGRs & $<2.5\times10^4$ & $>42.2$ & [8][9] \\
			SGRBs & $1.1\times10^3$ & $43.6$ & [6][10] \\
			LGRBs & $7\times10^2$ & $43.8$ & [6][11]\\
			\hline
		\end{tabular}
		\begin{tablenotes}
			\item (a) Event rate density in units of $\gpcyr$.
			\item (b) The threshold luminosity in FRB LF for the associated transient.
			\item (c) The references are: [1] \cite{Abb16b}; [2] \cite{Abbott17PRL}; [3] \cite{Map18}; [4] \cite{Thompson09arXiv}; [5] \cite{BM12ApJ}; [6] \cite{SZL15}; [7] \cite{Li11MNb}; [8] \cite{Ofek07ApJ} [9] \cite{Kulkarni14ApJ}; [10] \cite{CHP12}; [10] \cite{CTP07}.
		\end{tablenotes}
	\end{threeparttable}
	\label{tab:asso}
\end{table}

An immediate inference from \FIG{fig:ed} is that the FRB event rate density above
$10^{42}\,\ergs$ is comparable to that of supernovae, as known before \citep{Thornton13Sci}, but is greater than most compact star merger models. The WD-WD merger rate density is viable and is consistent with the Type Ia SN rate density. However, no SN has been observed to be associated with any FRB event, suggesting that a direct connection between FRBs and SN explosions can be ruled out. One may consider the scenarios that invoke once-in-a-life-time events from isolated neutron stars born from SN explosions, which may interpret the FRB event rate density. However, such genuinely non-repeating models proposed so far only apply to a sub-categories of SNe. For example, the ``blitzar'' model involving collapses of supramassive NSs to BHs \citep{FR14A&A} only applies to those SN explosions that form rapidly spinning, massive NSs whose gravitational mass exceeds the maximum mass of a non-spinning NS, which only comprise a small fraction of all SNe. Such a model may only account to a small fraction of FRBs (e.g. in the high-luminosity end). Another scenario invoking the product of SN explosions is the young magnetar model \citep{Murase16,Metzger17ApJ,Wang18}. This model also requires a special type of the progenitor system (e.g. those producing long GRBs or superluminous supernovae), and hence, is also only relevant for a small fraction of SN explosions. On the other hand, compared with the blitzar model, the young magnetar model can produce multiple bursts in its life time, making it possible to account for the majority of FRBs if most (if not all) of them are repeating sources \citep[e.g.][]{Ravi19}.

%to produce many bursts during their lifetimes, which is consistent with the existence of repeating FRB sources. In order not to over produce FRBs, one also requires that the SNe that make FRB engines should be only a small fraction of all SNe, e.g., related to those special progenitors that produce rapidly rotating, highly magnetized NSs within the young magnetar model }

Our results show that it is unlikely that all FRBs come 
from BH-BH, BH-NS or NS-NS mergers, \emph{if each merger event generates only 
one FRB}.
If such system can produce multiple FRBs, a substantial number of FRBs would be generated from each merger. The required multiplicity number of FRBs from each merger would be $10^{2}-10^{4}$ for BH-BH mergers, $10-100$ for NS-NS mergers, and $100-1000$ 
for NS-BH mergers. 
It can be difficult to produce many bursts associating with the merger event over the period of several years (as observed in FRB 121102). One possible scenario would be to invoke an NS-NS post-merger stable magnetar that can survive the merger \citep[e.g.][]{Z&M01ApJ,Metzger+08MN,Bucciantini+12MN,Rezzolla15ApJ,Margalit19,Wang20}. However, this requires a stiff NS equation of state and small masses in the NS-NS merger systems. Another possibility is that repeating FRBs are produced decades to centuries before the merger when the magnetospheres of the two NSs interact relentlessly \citep{Zhang20}.

On the other hand, our results do not object the possibility that a small fraction of FRBs, probably on the high-luminosity end, may be related to compact star mergers. \FIG{fig:ed} shows that for $L>10^{43}\,\ergs$, the merger models become viable. For example, an NS-NS merger may produce a bright FRB during the merger phase, which can be followed by weaker repeating bursts from a post-merger NS \citep{Jiang19}. 

It is interesting to note that the observed non-repeating FRBs typically have 
high luminosities with $L>10^{43}\,\ergs$, 
whereas the luminosity of the repeating bursts is usually smaller. 
This could account for the absence of repeaters detected by Parkes \citep{Petroff15b} and ASKAP \citep{James19MN} due to the sensitivity limits of the telescopes. One may suspect that the low-luminosity FRBs ($L<10^{43}\,\ergs$) could potentially be
the repeater candidates, because of their higher event rate and the existence of a plausible border in \FIG{fig:fdm}. With the same method in L18, we calculate the luminosity for individual FRB and predict a list of repeater candidates in the current FRBCAT, as shown in \TAB{tab:rpts}.

\begin{table}
    \centering
    \caption{The repeating FRB candidates according to the plausible luminosity clustering}
    \begin{threeparttable}
    \begin{tabular}{lccc}
    \hline\hline
       FRB & $\log(L_\mathrm{iso}/\ergs)$\tnote{a} & Telescope & Ref.\tnote{b} \\
       \hline
       010621 & $41.8^{+0.3}_{-0.7}$ & Parkes & [1][2] \\
       120127 & $42.9^{+0.3}_{-0.6}$ & Parkes & [3] \\
       140514 & $42.9^{+0.3}_{-0.6}$ & Parkes & [4] \\
       141113 & $40.7^{+0.3}_{-0.6}$ & Arecibo & [5] \\
       171020 & $42.5^{+0.3}_{-0.9}$ & ASKAP & [6] \\
       180301 & $42.4^{+0.4}_{-0.6}$ & Parkes & [7] \\
       180729.J1316+55 & $42.9^{+0.5}_{-0.8}$ & CHIME & [8] \\
       180923 & $42.3^{+0.3}_{-0.6}$ & Parkes & [9] \\
       \hline
    \end{tabular}
    \end{threeparttable}
    \begin{tablenotes}
    \footnotesize
    \item (a) The inferred isotropic luminosity within 95\% confidence interval.
    \item (b) The references are: [1] \cite{Keane11MN}; [2] \cite{Keane12MN}; [3] \cite{Thornton13Sci}; [4] \cite{Petroff15a}; [5] \cite{Patel18ApJ}; [6] \cite{Shannon18Nat}; [7] \cite{Price19MN}; [8] \cite{CHIME19Nat01}; [9] \cite{Bhandari18ATel}.
    \end{tablenotes}
    \label{tab:rpts}
\end{table}

\subsection{Searching plan}

The FRB detection rate as a function of minimum detection flux and field of view is given 
in \FIG{fig:map}. Obviously, telescopes with high gain and large field of view are the 
best for FRB searching, e.g.m CHIME and TianLai \citep{CHIME18ApJ, Chen12IJMPS}.  
Thanks to 13 beams receiver \citep{Staveley-Smith96PASA}, the Parkes 64-m radio 
telescope has played a pioneering role in FRB searching ever since its 
serendipitous discovery of the ``Lorimer Burst'' \citep{Lorimer07Sci}. 
Meanwhile, equipped with a 19-beam receiver \citep{Li18IMMag}, FAST can achieve 
a detection rate at the same level with Parkes telescope (see \FIG{fig:map}, 
about $5\times10^{-2}\,\mathrm{day}^{-1}$). This provides excellent prospects to 
detect many high-DM FRBs ($>3000\,\cmpc$), see also \cite{Zhang18bApJL}. Nevertheless, since the scattering effects is not included in the computation, 
the high gain telescopes (e.g., Arecibo and FAST) 
may detect less FRBs, if FRBs are in the scattering-limited regime.

For single-beam system, we calculated the detection rate - diameter relation as shown
in \FIG{fig:apert}. For the central value of LF, the optimal 
diameter of single-beam telescopes to detect FRBs is 30--40~m. Future discoveries made by CHIME, FAST, ASKAP and other instruments are probably crucial to our understanding for the FRB population. We anticipate that the methodologies developed in this paper and in L18 will form an important component in the analysis of these findings. 

\section*{Acknowledgements}
R.L., Y.P.M. and K.J.L. are supported by NSFC U15311243, XDB23010200, 2017YFA0402600 and 
funding from TianShanChuangXinTuanDui and Max-Planck Partner Group. W.Y.W. thanks the support of MoST grant 2016YFE0100300, the NSFC grants 11633004, NSFC 11473044, 11653003, and the CAS grants QYZDJ-SSW-SLH017, and CAS XDB 23040100. 
D.R.L. was supported by National Science Foundation award numbers AST-1516958 and OIA-1458952.
We thank Yuan-pei Yang, Ye Li and Xuelei Chen for helpful discussion and comments. Furthermore, we are grateful to the anonymous referee for the valuable criticisms and constructive suggestions. All the computations of Bayesian inference were performed on the 
cluster \textsc{dirac} at KIAA.
%%%%%%%%%%%%%%%%%%%%%%%%%%%%%%%%%%%%%%%%%%%%%%%%%%

%%%%%%%%%%%%%%%%%%%% REFERENCES %%%%%%%%%%%%%%%%%%

% The best way to enter references is to use BibTeX:

\bibliographystyle{mnras}
\bibliography{refs} % if your bibtex file is called example.bib

% Alternatively you could enter them by hand, like this:
% This method is tedious and prone to error if you have lots of references
%\begin{thebibliography}{99}
%\bibitem[\protect\citeauthoryear{Author}{2012}]{Author2012}
%Author A.~N., 2013, Journal of Improbable Astronomy, 1, 1
%\bibitem[\protect\citeauthoryear{Others}{2013}]{Others2013}
%Others S., 2012, Journal of Interesting Stuff, 17, 198
%\end{thebibliography}

%%%%%%%%%%%%%%%%%%%%%%%%%%%%%%%%%%%%%%%%%%%%%%%%%%

%%%%%%%%%%%%%%%%% APPENDICES %%%%%%%%%%%%%%%%%%%%%

\appendix
\section{Derivation of marginalised likelihood}
\label{app:margin}
More detailed calculation can be found in L18, here we just outline the major 
steps constructing the likelihood. 

The relations of isotropic luminosity - flux, intrinsic width - observed width and  
dispersion measures are
\begin{eqnarray}
	\log L  &=& \log S + 2\log d_{\rm L} + \log \Delta \nu_0 - \log \epsilon +\log 
	4\pi \, , \label{eq:rela1}\\
	\wi &=& \frac{\wo}{1+z} \, ,\label{eq:rela2}\\
	\DMh &=&(\DMe-\DMi) (1+z)-\DMs \, ,
	\label{eq:rela}
\end{eqnarray}
where $d_{\rm L}$ is the luminosity distance, $\Delta \nu_0$ is the spectrum 
width. Note that in this paper, we use a flat spectrum as the FRB spectrum and fix the 
spectrum width at a reference value $\Delta \nu_0=1\,\rm GHz$. $\DMi$ is the DM 
contributed by intergalactic medium (IGM), given by \citep{DZ14ApJ}
\begin{equation}
	 \DMi(z)=\frac{\rho_{\rm c} \Omega_{\rm b} f_{\rm IGM}}{m_{\rm p} H_0}\int \frac{g(z)(1+z)}{E(z)}\,\D z\,,
\label{eq:dmigm}
\end{equation}
where $m_{\rm p}$ is the proton mass, $H_0$ is the Hubble constant. $\rho_c$ is the current critical density of the universe, $\Omega_b$ is the mass fraction in the universe. The $f_{\mathrm{IGM}}$ is the cosmological baryon mass fraction in the IGM, here we use $f_{\mathrm{IGM}}\simeq0.83$  \citep{Fukugita98ApJ}. The function $g(z)$ is the ionized electron number fraction per baryon, given as
\begin{equation}
g(z)\simeq\frac{3}{4}\chi_{e,\mathrm{H}}(z)+\frac{1}{8}\chi_{e,\mathrm{He}}(z)\,,
\end{equation}
where $\chi_{e,\mathrm{H}}$ and $\chi_{e,\mathrm{He}}$ are the cosmic ionization fraction of hydrogen and helium, respectively.

Based on \EQ{eq:rela1}, (\ref{eq:rela2}) and (\ref{eq:rela}),  we can convert 
the PDF $f(\log L, \log \wi, N, z, \DMh, \DMs, \log\epsilon)$ to $f(\log S, \log 
\wo, N, \DMe, z, \DMs, \log\epsilon)$ using the Jacobian transformation, i.e.
\begin{equation}
\begin{aligned}
	& f(\log S, \log \wo, N, \DMe, z, \DMs, \log\epsilon) \\
	={} & |\MX{J}|\, f(\log L, \log \wi, N, \DMh, z, \DMs, \log\epsilon)\,,
\end{aligned}
\end{equation}
where the Jacobian determinant is written as
\begin{equation*}
\begin{split}
	\left|\MX{J}\right| &= \left| \frac{\pt (\log L,\,\log \wi,\,N,\,\DMh,\,z,\,\DMs,\,\log\epsilon)}{\pt (\log S,\, \log \wo,\, N,\,\DMe,\,z,\,\DMs,\,\log\epsilon)} \right|
	\label{eq:jocdet} \\
	&=\left|\begin{pmatrix}
		1 & 0 & 0 & 0 & \pt\log L/\pt z & 0 & -1 \\
		0 & 1 & 0 & 0 & \wo/(\ln10\,\wi(1+z)^2) & 0 & 0 \\
		0 & 0 & 1 & 0 & 0 & 0 & 0 \\
		0 & 0 & 0 & 1+z & \pt\DMh/\pt z & -1 & 0 \\
		0 & 0 & 0 & 0 & 1 & 0 & 0 \\
		0 & 0 & 0 & 0 & 0 & 1 & 0 \\
		0 & 0 & 0 & 0 & 0 & 0 & 1 \\
\end{pmatrix}\right| \\
&= 1+z.
\end{split}
\end{equation*}
One gets
\begin{equation}
\begin{aligned}
	& f(\log S, \log \wo, N, \DMe, z, \DMs, \log\epsilon) \\
	&={} \phi(\log L)\, \fw(\log \wi)\, P(N)\, \fd(\DMh|z)\, \fz(z)\, \fs(\DMs) \\
	& \quad \cdot \fe(\log\epsilon)\,(1+z) \,,
\end{aligned}
\end{equation}
Adding the likelihood for the number of events to above equation, we have
\begin{equation}
\begin{aligned}
	& f(\log S, \log \wo, N, \DMe, z, \DMs, \log\epsilon) \\
	& ={}  P(N)\cdot f(\log S, \log \wo, \DMe, z, \DMs, \log\epsilon)\,,
\end{aligned}
\end{equation}
We marginalise the $f(\log S, \log \wo, \DMe, z, \DMs, \log\epsilon)$ by 
integrating $z, \DMs$ and $\log\epsilon$.
\begin{equation}
\begin{aligned}
	f(\log S, \log \wo, \DMe)&=\frac{1}{\nf} \int_0^{\infty}
	I(\log L)\,\fw(\log \wi)\fz(z)\, \\ 
	&\quad I(\DMe, z) \, (1+z)\, \D z ,
\end{aligned}
	\label{eq:likfun}
\end{equation}
The marginalisation of $\DMs$ leads to
\begin{equation}
\begin{aligned}
I(\DMe, z) &= \int_0^{\max(\DMs)}\fd(\DMh|z) \fs(\DMs)\, \D\DMs \, ,
\end{aligned}
\end{equation}
and the marginalisation of $\log\epsilon$ gives
\begin{equation}
\begin{aligned}
	I(\log L)=\int \phi(\log L) \fe(\log\epsilon)\, \D\log\epsilon.
\end{aligned}
\end{equation}
The normalisation factor for $f(\log S, \log \wo, \DMe, z)$ becomes
\begin{equation}
\begin{aligned} 
    \nf &= \int_{\log S_{\rm min}(\wo)}^{\infty}\,\D \log S \int \int \int f(\log S, \log \wo, \DMe, z)\, \\
     & \quad \cdot \D \log \wo \, \D \DMe \, \D z.  \\ 
\end{aligned}
\end{equation}
The lower limit of the flux density integration, $S_{\rm min}$, is the minimum detectable flux density for the telescope at the time when a given FRB with pulse width $\wo$ was detected, which is mentioned in \EQ{eq:smin} in the main text.

The final marginalised likelihood function is
\begin{equation}
	\mathcal{L}=\mathcal{L}(N)\cdot\mathcal{L}(\log S, \log \wo, \DMe)\,,
\end{equation}
where
\begin{equation}
\begin{aligned}
	\mathcal{L}(\log S, \log \wo, \DMe)&=\frac{1}{\nf} \int_0^{\infty} I(\log L)\, \fw(\log \wi)\, \fz(z)\, \\ 
	&\quad \cdot I(\DMe, z) \, (1+z)\,\D z ,
\end{aligned}
\end{equation}
and
\begin{equation}
	\mathcal{L}(N)=\frac{(\rho \Omega t)^N e^{-\rho \Omega t}}{N!}\,.
\end{equation}

Compared to the method of L18, we add the modelling for the number of events, 
which is the key to the even rate density inference.

\section{Derivation on the surface event rate of survey}
\label{app:rho}
For the $k$-th FRB in the $j$-th survey, the surface event rate is defined as the partial derivative of detection number $N$ to observing time $t_{\rm obs}$ and FoV $\Omega$,
\begin{equation}
\begin{aligned}
\rho_{kj} &=\frac{\pt N(>S_{\mathrm{min},\, kj})}{\pt t_{\mathrm{obs}} \pt \Omega} \\
&= \int \frac{\pt V}{\pt \Omega\pt z} \D z \int_{L_{\mathrm{min},\, kj}}^{\infty}\frac{\pt N}{\pt V\pt t_{\mathrm{obs}}\, \pt\log L}\D\log L \\
& = \int \frac{\D t}{\D t_\mathrm{obs}}\frac{\pt V}{\pt \Omega\pt z} \D z \int_{L_{\mathrm{min},\, kj}}^{\infty} \D \log L\, \int_{\log\frac{1}{2}}^{0} \frac{\pt N}{\pt V\, \pt t\, \pt\log L\, \pt\log\epsilon} \\
&\quad \cdot \D\log\epsilon\,  \\
& = \int_0^{\infty} \frac{1}{1+z}\frac{r(z)^2}{H(z)}\D z  \int_{\log L_{\mathrm{min}},\, kj}^{\infty} I(\log L)\,\D\log L \,.
\end{aligned}
\label{eq:deri}
\end{equation}
The lower limit of above luminosity integral is determined by $L_{\mathrm{min},\, kj}=\max(L_0, L_{\mathrm{thre},\, kj})$, where $L_0$ is the intrinsic lower cut-off of LF and threshold luminosity of survey $L_\mathrm{thre}$ is given as 
\begin{equation}
	L_{\mathrm{thre},\, kj}\equiv4\pi d_{\rm L}^2 \Delta\nu_0 S_{\mathrm{min},\, kj}\,,
\end{equation}
where the flux threshold of FRB with a duration $w_{\mathrm{o},\,kj}$ is 
\begin{equation}
	S_{\mathrm{min}, kj}=\frac{\mathrm{S/N}_0\ T_{\mathrm{sys}}}{G\sqrt{N_\mathrm{p}\,\mathrm{BW}\, w_{\mathrm{o},\,kj}}}\,.
\end{equation}

Alternatively, if we marginalise the luminosity in advance, the surface event rate in \EQ{eq:deri} would be re-written as
\begin{equation}
\begin{aligned}
	\rho_{kj} &= \int_0^{\infty} \frac{1}{1+z}\frac{r(z)^2}{H(z)} \D z\, \int_{\log\frac{1}{2}}^{0} \phi^* \Gamma\left(\alpha+1, \frac{L_{\mathrm{min},\,kj}}{\epsilon L^*}\right)\fe(\log\epsilon) \\
	&\quad \cdot \D\log\epsilon \, ,
\end{aligned}
\end{equation}
where $\Gamma$ is the incomplete \textsc{gamma} function.
Due to the FRB width distribution $\fw(\log \wo)$, the total FRB surface rate 
becomes
\begin{equation}
\rho_j=\int \rho(\wo)\, \fw(\log \wo)\, \D \log \wo\,.
\end{equation}

Overall, one will have the surface rate of $j$-th survey by marginalising the 
FRB intrinsic widths, \begin{equation}
\begin{aligned}
	\rho_j &= \int_0^{\infty} \frac{1}{1+z}\frac{r(z)^2}{H(z)}\D z \int \fw(\log \wi)\,\D\log \wi \\
	&\quad \cdot \int_{\log L_{\mathrm{min}}(\wo)}^{\infty} I(\log L)\,\D\log L,
\end{aligned}
\end{equation}
where $I(\log L)=\int \phi(\log L) \fe(\log\epsilon)\,\D\log\epsilon$.

%%%%%%%%%%%%%%%%%%%%%%%%%%%%%%%%%%%%%%%%%%%%%%%%%%

\section{Adding independent Poisson processes}
\label{app:mpoi}
Denote $\lambda_1$ and $\lambda_2$ as the rates of two independent event $X$ and 
$Y$, respectively. We now compute the event rate of detecting either one of 
them. If we divide the total observing time $t$ into $n$ pieces, the 
probability of detecting $k$ events in all $n$ durations is
\begin{equation}
	\begin{aligned}
		P(X+Y=k)&=\binom{n}{k} \sum_{i=0}^k \left[\binom{k}{i}(\lambda_1\Delta 
		t)^i(\lambda_2\Delta t)^{k-i}(1-\lambda_1\Delta t)^{n-i}\right. \\
		& \quad \left.\cdot (1-\lambda_2\Delta t)^{n-k+i}\right] \\
		&=\binom{n}{k} \sum_{i=0}^k \left[\binom{k}{i}\left(\frac{\lambda_1t}{n}\right)^i\left(\frac{\lambda_2t}{n}\right)^{k-i}\left(1-\frac{\lambda_1t}{n}\right)^{n-i} \right. \\
		& \quad \left.\cdot \left(1-\frac{\lambda_2t}{n}\right)^{n-k+i}\right]
	\end{aligned}
\end{equation}
Using the relation of
\begin{equation}
	\lim_{n\to\infty}\left(1-\frac{\lambda t}{n}\right)^n = e^{-\lambda t}\,,
\end{equation}
one has \begin{equation}
	\begin{aligned}
		\lim_{n\to\infty}P(X+Y=k)&=\frac{n!}{(n-k)!k!n^k}\sum_{i=0}^k\left[\frac{k!}{(k-i)!i!}(\lambda_1t)^i(\lambda_2t)^{k-i} \right.\\
		& \quad \left.\cdot e^{-\lambda_1t}e^{-\lambda_2t} \right] \\
		&=\frac{e^{-(\lambda_1+\lambda_2)t}}{k!}\sum_{i=0}^k\left[\binom{k}{i}(\lambda_1t)^i(\lambda_2t)^{k-i}\right] \\
		&=\frac{[(\lambda_1+\lambda_2)t]^ke^{-(\lambda_1+\lambda_2)t}}{k!}\,.
	\end{aligned}
\end{equation} This is the continuous time limit. Thus, the event rate is 
additive and the distribution for the number of detection follows Poisson 
distribution, if each event is independent. In a similar fashion, we can show 
that for $N$ random events \{${X_1, X_2, \cdots, X_N}$\}, the total rate of 
detecting any of them is
$\lambda=\lambda_1+\lambda_2+\cdots+\lambda_N$.

\section{Algorithm verification using the mock data}
\label{app:post}
\begin{figure*}
\centering
\subfloat[Sample I]
{\includegraphics[width=0.8\textwidth]{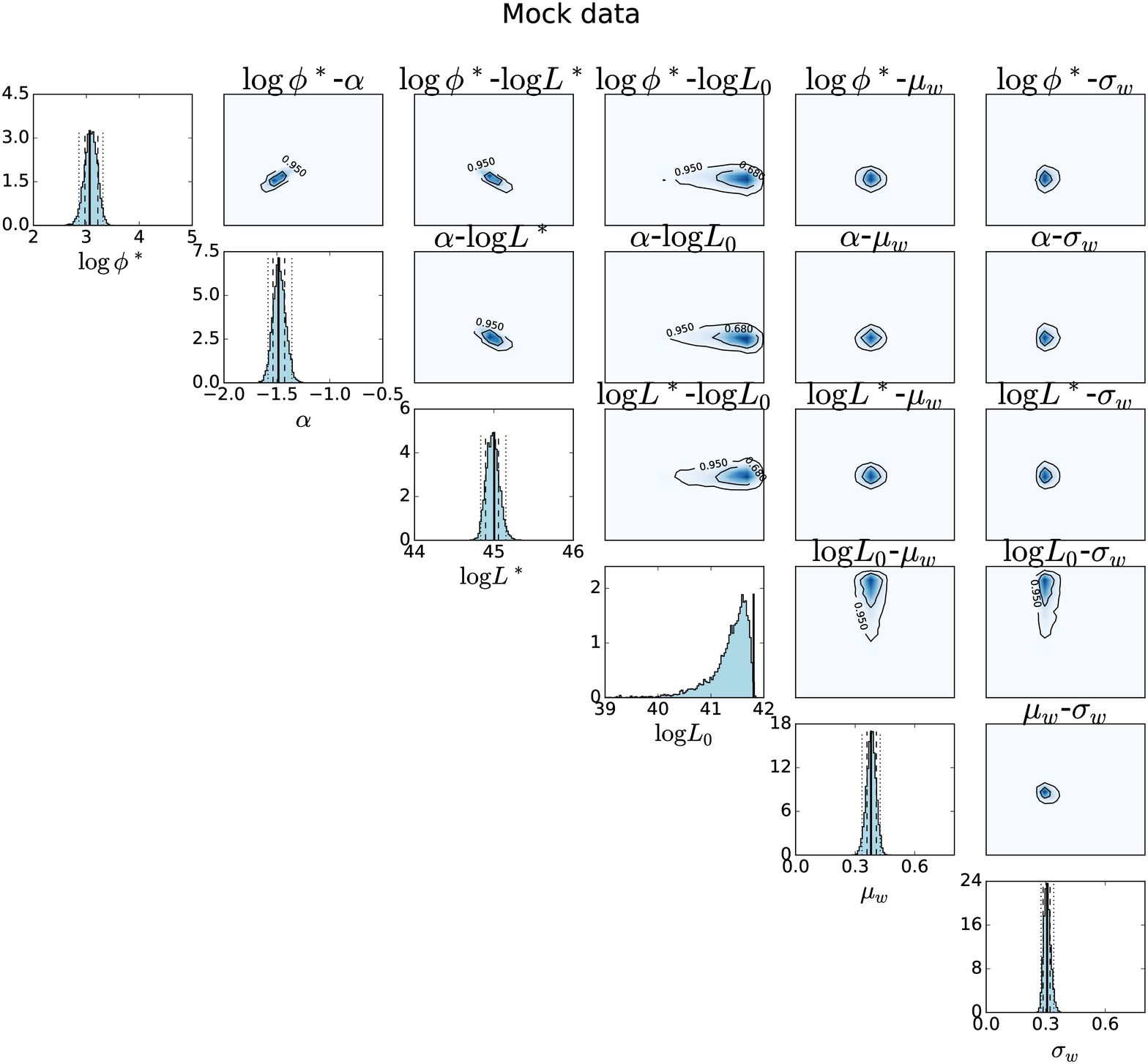}} \\
\subfloat[Sample II]
{\includegraphics[width=0.8\textwidth]{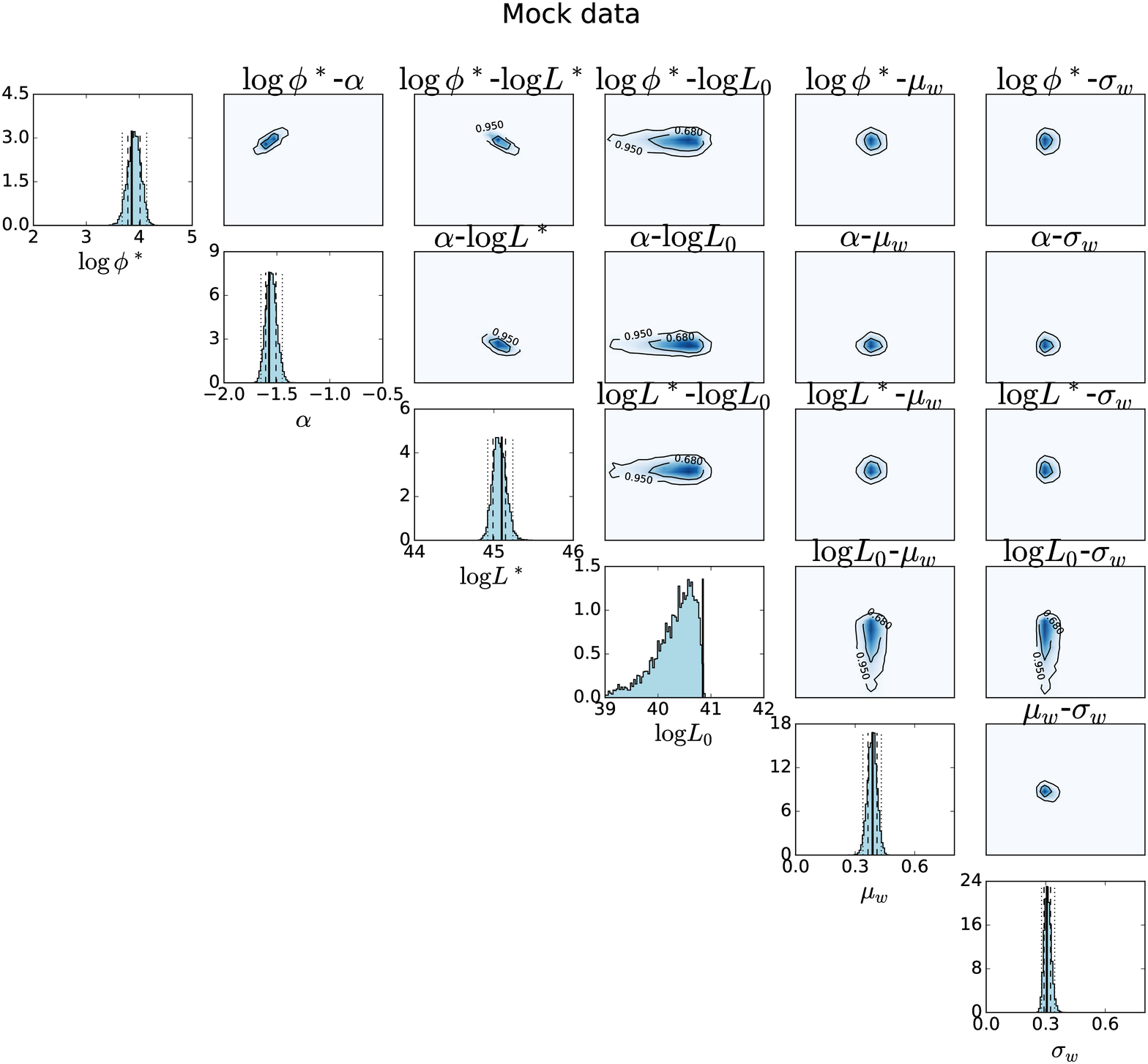}}
\caption{Posterior distributions of the inferred parameters of mock FRB samples (I and II). In each panel, subplots are the same as \FIG{fig:post}.
%The diagonal histograms are the marginalised one-dimensional posterior distributions for all the parameters. For the parameters except $\log L_0$, the solid lines denote the most probable measured values, the dashed lines and dotted lines denote 68\% and 95\% confidence intervals respectively. For $\log L_0$, we can only determine the upper limit, where the solid line is for 95\% C.L. The off-diagonal contour plots are for the marginalised two-dimensional posteriors, with parameters indicated in the plot titles. The inner and outer black contours are for 68\% and 95\% C.L. respectively.
}
\label{fig:post_mock}
\end{figure*}

For the simulations, the information of the two mock surveys are described in \TAB{tab:mocsvy}. After Bayesian inference, the posterior distribution of mock FRBs are shown in \FIG{fig:post_mock}, and the 
results of parameter inference are listed in \TAB{tab:mocres}. As one can see, 
our algorithm correctly recovered parameter central value used in the simulations with 95\% confidence.

\begin{table*}
\caption{Systematic information of the mock surveys}
\centering
\begin{tabular}{cccccccc}
\hline
\hline
Survey & G & $T_\mathrm{sys}$ & BW & $\Omega$ & $N_\mathrm{p}$ & S/N$_{0}$ & $N_\mathrm{FRB}$\\
& (K/Jy) & (K) & (MHz) & ($\sqdeg$) & & & \\
\hline
S1 & 0.7 & 30 & 300 & 0.55 & 2 & 10 & 100 \\
S2 & 0.05 & 100 & 300 & 30 & 2 & 10 & 100 \\
\hline
\end{tabular}
\label{tab:mocsvy}
\end{table*}

\begin{table*}
\caption{The results of Bayesian inference for the mock FRBs}
\centering
\begin{threeparttable}
\begin{tabular}{ccccccccccccccc}
\hline
\hline
\multirow{2}[3]{*}{Sample} & \multicolumn{2}{c}{$t_{\rm svy}$\,(hr)} & \multicolumn{2}{c}{$\phi^*\,(\gpcyr)$} & \multicolumn{2}{c}{$\alpha$} & \multicolumn{2}{c}{$\log L^*\,(\ergs)$} & \multicolumn{2}{c}{$\log L_0\,(\ergs)$} & \multicolumn{2}{c}{$\mu_w$} & \multicolumn{2}{c}{$\sigma_w$} \\
\cmidrule(lr){2-3} \cmidrule(lr){4-5} \cmidrule(lr){6-7} \cmidrule(lr){8-9} \cmidrule(lr){10-11} \cmidrule(lr){12-13} \cmidrule(lr){14-15} & S1 & S2 & Fid.\tnote{(a)} & Mea.\tnote{(b)} & Fid. & Mea. & Fid. & Mea. & Fid. & Mea. & Fid. & Mea. & Def. & Mea. \\
\midrule
I & $15149$ & $5366$ & $1\times10^3$ & $1148_{-424}^{+894}$ & $-1.5$ & $-1.48_{-0.10}^{+0.13}$ & 45.0 & $45.00_{-0.17}^{+0.15}$ & 40.0 & $\le41.8$ & 0.4 & $0.38_{-0.05}^{+0.05}$ & 0.3 & $0.31_{-0.03}^{+0.03}$ \vspace{0.5em} \\
II & $1383$ & $647$ & $1\times10^4$ & $7244_{-2458}^{+6559}$ & $-1.5$ & $-1.57_{-0.08}^{+0.13}$ & 45.0 & $45.10_{-0.18}^{+0.14}$ & 40.0 & $\le40.8$ & 0.4 & $0.39_{-0.05}^{+0.04}$ & 0.3 & $0.31_{-0.03}^{+0.04}$ \\
\bottomrule
\end{tabular}
\begin{tablenotes}
\footnotesize
\item (a) Fiducial values when we simulate the mock data.
\item (b) All the measurements are given within 2-$\sigma$ error.
\end{tablenotes}
\end{threeparttable}
\label{tab:mocres}
\end{table*}

% Don't change these lines
\bsp	% typesetting comment
\label{lastpage}
\end{document}